\documentclass[11pt]{article}
\usepackage{amssymb}

\usepackage{makeidx}
\usepackage[totalwidth=460truept,totalheight=610truept]{geometry}
\usepackage{latexsym,amssymb,amsmath,graphicx,accents,eucal,slashed,subfigure}
\usepackage{dsfont}
\usepackage[T1]{fontenc}
\usepackage{hyperref}

\global\arraycolsep=1truept

\numberwithin{equation}{section}

\begin{document}

\bigskip \hfill IFUP-TH 2012/08

\vskip 1.4truecm

\begin{center}
{\huge \textbf{A General Field-Covariant}}

\vskip .5truecm

{\huge \textbf{Formulation Of Quantum Field Theory}}

\vskip 1truecm

\textsl{Damiano Anselmi} \vskip .2truecm

\textit{Dipartimento di Fisica ``Enrico Fermi'', Universit\`{a} di Pisa, }

\textit{Largo B. Pontecorvo 3, I-56127 Pisa, Italy,}

\vskip .2truecm

damiano.anselmi@df.unipi.it

\vskip 1.5truecm

\textbf{Abstract}
\end{center}

\medskip

{\small In all nontrivial cases renormalization, as it is usually
formulated, is not a change of integration variables in the functional
integral, plus parameter redefinitions, but a set of replacements, of
actions and/or field variables and parameters. Because of this, we cannot
write simple identities relating bare and renormalized generating
functionals, or generating functionals before and after nonlinear changes of
field variables. In this paper we investigate this issue and work out a
general field-covariant approach to quantum field theory, which allows us to
treat all perturbative changes of field variables, including the relation
between bare and renormalized fields, as true changes of variables in the
functional integral, under which the functionals }$Z${\small \ and }$W=\ln Z$%
{\small \ behave as scalars. We investigate the relation between composite
fields and changes of field variables, and we show that, if }$J${\small \
are the sources coupled to the elementary fields, all changes of field
variables can be expressed as }$J${\small -dependent redefinitions of the
sources }$L${\small \ coupled to the composite fields. We also work out the
relation between the renormalization of variable-changes and the
renormalization of composite fields. Using our transformation rules it is
possible to derive the renormalization of a theory in a new variable frame
from the renormalization in the old variable frame, without having to
calculate it anew. We define several approaches, useful for different
purposes, in particular a linear approach where all variable changes are
described as linear source redefinitions. We include a number of explicit
examples. }

\vskip 1truecm

\vfill\eject

\section{Introduction}

\setcounter{equation}{0}

The present formulation of quantum field theory is not sufficiently general.
Several properties we are interested in depend on the variables we use to
formulate and quantize the theory. For example, power-counting
renormalizability requires that the action should contain no parameters of
negative dimensions in units of mass, but this property is spoiled by a
general change of field variables. If we work in a generic field-variable
setting, the only way we have to state the power-counting criterion is to
demand that there should exist a field-variable frame where the theory
becomes renormalizable according to the usual rules. We do not have a
field-covariant formulation of quantum field theory, and we lack efficient
variable-independent criteria to identify theories belonging to special
classes, such as the renormalizable, conformal and finite theories. We can
state, for example, that a theory is finite if all divergences can be
reabsorbed by means of field redefinitions, but this is just the definition
of finite theory, not a criterion to identify finite theories. So far the
only general criterion we seem to have is ``calculate and see'', which is
clearly unsatisfactory. Similarly, we can define renormalizable theories as
those whose divergences can be subtracted by means of field redefinitions
and redefinitions of a finite number of physical parameters. Yet, this is
not an efficient criterion to identify them. We are thus stuck with power
counting and other criteria tied to special field-variable frames, and miss
a broader view and a deeper insight.

The first task to overcome these difficulties is to develop a general
field-covariant approach. Once this result is obtained, the second problem
is to work out criteria that allow us to identify theories belonging to
special subclasses in the most economic and efficient way. In this paper we
investigate the first issue. The best way to search for a field-covariant
formulation of quantum field theory is to study the most general
perturbative changes of field variables. If we do this, we realize that
whenever renormalization involves nonlinear field redefinitions the usual
relation between bare and renormalized fields is not a true change of
variables in the functional integral, but just a ``replacement''. This means
that the action is transformed according to the field redefinition, but the
term $\int J\varphi $, which identifies the ``elementary field'' used to
write Feynman rules and calculate diagrams, is not transformed, but just
replaced with $\int J^{\prime }\varphi ^{\prime }$, the analogous term in
the new variables. These operations do give the transformed generating
functional, but spoil its relation with the starting generating functional.
Replacements are enough for a number of purposes, but they are not
satisfactory for a covariant approach to quantum field theory.

Consider for example the infinitesimal redefinition $\varphi =\varphi
^{\prime }+b\varphi ^{\prime \hspace{0.01in}2}$, with $b\ll 1$. When we use
it as a change of variables in the functional integral, the term $\int
J\varphi $ is turned into $\int J\varphi ^{\prime }+b\int J\varphi ^{\prime 
\hspace{0.01in}2}$ and the transformed functional integral is no longer
written in the usual way, by which we mean that it does not depend on $J$
only through the term $\int J\varphi ^{\prime }$. To solve this problem we
have the freedom to define a suitable $J^{\prime }$, or correct the field
redefinition. However, correcting the field redefinition is not helpful,
since it can at most generate terms proportional to the field equations, to
the lowest order. Thus, we must search for a $J^{\prime }$ such that $\int
J\varphi =\int J^{\prime }\varphi ^{\prime }$. To the first order in $b$ we
find $J=J^{\prime }-b\varphi ^{\prime }J^{\prime }$. This $J$-redefinition
is not acceptable, because $J$ is an external source and $\varphi $ is an
integrated field. Thus, normally we jump to the new generating functional
replacing $\int J\varphi $ with $\int J^{\prime }\varphi ^{\prime }$ by
brute force. In this paper we show how to overcome these problems and
promote every manipulation to a true change of integration variables in the
functional integral.

We show that it is always possible to reformulate the map relating bare and
renormalized quantities, which we call \textit{BR\ map}, as a true change of
variables, plus redefinitions of parameters. Then, we study the most general
perturbative changes of field variables and show how they get reflected from
the classical action to the functionals $Z$ and $W=\ln Z$. The $\Gamma $%
-functional requires a separate investigation, so in this paper we
concentrate our attention on the $Z$- and $W$-functionals.

There is an intrinsic relation between composite fields, often called
``composite operators'', and changes of field variables. Indeed, any local
nonlinear change of variables maps elementary fields into composite ones.
However, the field that we call elementary and with respect to which we
perform the quantization is just a personal choice among infinitely many.
Since the physics cannot depend on the variables we use, there are no
intrinsic notions of ``elementary fields'' and ``composite fields''. For
this reason, it is convenient to treat the theory together with the set of
its composite fields. Changes of field variables undergo their own
renormalization, which is related to the renormalization of composite fields.

A perturbative field redefinition is a field redefinition that can be
expressed as the identity map plus a perturbative series of local monomials
of the fields and their derivatives. We show that, if $J$ denote the sources
coupled to the elementary fields, the most general perturbative change of
field variables is a $J$-dependent redefinition of the sources coupled to
the composite fields, and the $Z$- and $W$-functionals behave as scalars.
Taking advantage of these properties, we can easily relate correlation
functions before and after the variable change. In particular, our results
provide a simple method to derive the renormalization of the theory in the
new variables from the renormalization of the theory in the old variables,
without having to calculate it anew.

We use several approaches and compare their virtues and weaknesses. In one
approach, which we call \textit{redundant}, descendants and composite fields
proportional to the field equations are treated as independent composite
fields. In another approach, called \textit{essential}, they are not
considered independent, therefore suppressed. In a third approach, called 
\textit{linear}, we are able to linearize the source redefinitions that
encode the most general changes of field variables.

For definiteness, we work using the Euclidean notation and the
dimensional-regularization technique, but no results depend on these
choices. To simplify the presentation, we imagine that the fields we are
working with are bosonic, so we do not need to pay attention to their
positions. The arguments can be immediately generalized to include fermionic
fields and Grassmann variables.

The paper is organized as follows. In section 2 we address the problem. In
section 3 we define the various approaches we are going to use. In sections
4--7 we study the BR map in each approach. In sections 8--10 we study the
most general changes of field variables, in each approach, at the bare and
renormalized levels. In section 11 we make some important remarks about the
relation between bare and renormalized changes of field variables. We give a
number of explicit examples in section 12. Section 13 contains our
conclusions.

\section{Description of the problem}

\setcounter{equation}{0}

Normally we formulate quantum field theory starting from a classical action $%
S_{c}(\varphi ,\lambda )$, where $\varphi $ is the set of fields and $%
\lambda $ is the set of parameters (including both couplings and masses). We
define generating functionals, try to calculate them perturbatively, and
find divergences. We discover that at every step of the perturbative
subtraction divergences are local, and therefore can be removed with
redefinitions of fields and parameters in the classical action, provided the
classical action contains enough independent parameters. When it is not so,
we just introduce new parameters at the tree level.

The subtraction of divergences is done as follows. Every quantity has a bare
version, which is basically the classical version, and a renormalized
version. The renormalized fields and parameters are those that make the
generating functionals convergent. We call any map relating bare and
renormalized quantities \textit{BR map}. The subtraction of divergences
makes the renormalized quantities depend on one parameter more than the bare
quantities, the ``dynamical scale'' $\mu $, so the BR map has the form 
\begin{equation}
\varphi _{\mathrm{B}}=\varphi _{\mathrm{B}}(\varphi ,\lambda ,\mu ),\qquad
\lambda _{\mathrm{B}}=\lambda _{\mathrm{B}}(\lambda ,\mu ).  \label{BRrel}
\end{equation}
The relations $\varphi _{\mathrm{B}}=\varphi _{\mathrm{B}}(\varphi ,\lambda
,\mu )$ are perturbatively local, but need not be polynomial. For example,
they are not polynomial in a generic non-renormalizable theory, such as
Einstein gravity \cite{thooftveltman}.

The bare action $S_{\mathrm{B}}(\varphi _{\mathrm{B}},\lambda _{\mathrm{B}})$
coincides with the classical action $S_{c}(\varphi ,\lambda )$, once fields
and parameters are replaced with the bare ones: $S_{\mathrm{B}}(\varphi _{%
\mathrm{B}},\lambda _{\mathrm{B}})=S_{c}(\varphi _{\mathrm{B}},\lambda _{%
\mathrm{B}})$. Similarly, the renormalized action coincides with the bare
action once bare fields and parameters are expressed in terms of the
renormalized ones: 
\begin{equation}
S(\varphi ,\lambda ,\mu )=S_{\mathrm{B}}(\varphi _{\mathrm{B}},\lambda _{%
\mathrm{B}}).  \label{amalga}
\end{equation}

In the Euclidean notation the bare and renormalized $Z$- and $W$-generating
functionals are 
\begin{eqnarray}
Z_{\mathrm{B}}(J_{\mathrm{B}},\lambda _{\mathrm{B}}) &=&\int [\mathrm{d}%
\varphi _{\mathrm{B}}]\hspace{0.02in}\exp \left( -S_{\mathrm{B}}(\varphi _{%
\mathrm{B}},\lambda _{\mathrm{B}})+\int \varphi _{\mathrm{B}}J_{\mathrm{B}%
}\right) =\exp \left( W_{\mathrm{B}}(J_{\mathrm{B}},\lambda _{\mathrm{B}%
})\right) ,  \nonumber \\
Z(J,\lambda ,\mu ) &=&\int [\mathrm{d}\varphi ]\hspace{0.02in}\exp \left(
-S(\varphi ,\lambda ,\mu )+\int \varphi J\right) =\exp \left( W(J,\lambda
,\mu )\right) ,  \label{ZBJL}
\end{eqnarray}
and the $\Gamma $-functionals $\Gamma _{\mathrm{B}}(\Phi _{\mathrm{B}%
},\lambda _{\mathrm{B}})$ and $\Gamma (\Phi ,\lambda ,\mu )$ are the
Legendre transforms of $W_{\mathrm{B}}$ and $W$ with respect to $J_{\mathrm{B%
}}$ and $J$, respectively.

Our investigation starts from the following problem: \textit{are there
relations } 
\[
J_{\mathrm{B}}=J_{\mathrm{B}}(J,\lambda ,\mu ),\qquad \Phi _{\mathrm{B}%
}=\Phi _{\mathrm{B}}(\Phi ,\lambda ,\mu ), 
\]
\textit{such that } 
\begin{equation}
Z_{\mathrm{B}}(J_{\mathrm{B}},\lambda _{\mathrm{B}})=Z(J,\lambda ,\mu
),\quad W_{\mathrm{B}}(J_{\mathrm{B}},\lambda _{\mathrm{B}})=W(J,\lambda
,\mu ),\quad \Gamma _{\mathrm{B}}(\Phi _{\mathrm{B}},\lambda _{\mathrm{B}%
})=\Gamma (\Phi ,\lambda ,\mu )~?  \label{question}
\end{equation}

If the renormalization of $\varphi $ is multiplicative, such relations exist
and are easy to find. We have 
\begin{equation}
\varphi _{\mathrm{B}}=Z_{\varphi }^{1/2}\varphi ,\qquad J_{\mathrm{B}%
}=Z_{\varphi }^{-1/2}J,\qquad \Phi _{\mathrm{B}}=Z_{\varphi }^{1/2}\Phi
_{{}},  \label{bu}
\end{equation}
where $Z_{\varphi }$ is the wave-function renormalization constant of the
elementary field $\varphi $. The second relation of (\ref{bu}) is obtained
applying the change of variables $\varphi _{\mathrm{B}}=Z_{\varphi
}^{1/2}\varphi $ in the functional integral that defines $Z_{\mathrm{B}}(J_{%
\mathrm{B}},\lambda _{\mathrm{B}})$. This operation gives indeed $%
Z(J,\lambda ,\mu )$ once we define $J_{\mathrm{B}}=Z_{\varphi }^{-1/2}J$.
The third relation is obtained applying the same change of variables to go
from $\Phi _{\mathrm{B}}=\langle \varphi _{\mathrm{B}}\rangle $ to $\Phi
=\langle \varphi \rangle $.

Nevertheless, when the relation between $\varphi _{\mathrm{B}}$ and $\varphi 
$ is not multiplicative, the matter is more complicated. Let us make the
change of variables $\varphi _{\mathrm{B}}=\varphi _{\mathrm{B}}(\varphi
,\lambda ,\mu )$ in $Z_{\mathrm{B}}(J_{\mathrm{B}},\lambda _{\mathrm{B}})$
again. Since the function $\varphi _{\mathrm{B}}(\varphi ,\lambda ,\mu )$ is
perturbatively local, using the dimensional-regularization technique the
Jacobian determinant is identically 1 (because every polynomial of the
momenta integrates to zero), so the functional integration measure is
invariant under the BR map. The action in the exponent transforms correctly,
because of (\ref{amalga}), but there is no obvious way to transform the term 
$\int \varphi _{\mathrm{B}}J_{\mathrm{B}}$ into $\int \varphi J$. Thus, we 
\textit{cannot} conclude $Z_{\mathrm{B}}(J_{\mathrm{B}},\lambda _{\mathrm{B}%
})=Z(J,\lambda ,\mu )$.

Formula (\ref{ZBJL}) is a very specific way to express a functional
integral, which does not survive a generic change of field variables. The
entire $J$-dependence is encoded in the term $\int \varphi J$ appearing in
the exponent of the integrand. We say that, in this the case, the generating
functionals $Z$ and $W$ are written in the \textit{conventional} \textit{form%
}. The role of $\int \varphi J$ is to specify which is the elementary field
used to derive the Feynman rules and calculate diagrams. Clearly, the
elementary field is spoiled by a nonlinear change of field variables.

A generic change of variables, including a translation $\varphi \rightarrow
\varphi +a$, converts the functional integral to some unconventional form.
We can also make perturbatively local non-polynomial changes of variables
that depend on $J$. What is not obvious is how to go back to the
conventional form after the variable-change. In section 9 we give a theorem
that allows us to achieve this goal, in the most general perturbative
setting.

To answer the question raised above we must first introduce composite fields
and study their renormalization. The reason is that nonlinear field
redefinitions always mix elementary fields and composite fields. The
renormalization of composite fields has been extensively treated in the
literature \cite{compo}, but we must revisit it before proceeding. Actually,
it is necessary to formulate a number of different approaches, because each
of them is convenient for a different purpose.

Once this is done, we can study the BR map in a general setting. We
basically have two options to describe the field redefinitions contained in
BR\ maps. One option is to make a classical change of variables inside the
action and replace the bare term $\int J_{\mathrm{B}}\varphi _{\mathrm{B}}$
with the renormalized one $\int J\varphi $ by brute force. This is the
operation we are accustomed to, and we call it a \textit{replacement}. The
other option is to make a true change of variables inside the functional
integral, which is the new operation we investigate in this paper.

Instead of working with generating functionals we could just work with sets
of correlation functions $\langle O_{1}(\varphi (x_{1}))\cdots O_{n}(\varphi
(x_{n}))\rangle $ at $J=0$, because they are manifestly invariant under
changes of field variables $\varphi ^{\prime }=\varphi ^{\prime }(\varphi )$%
. Indeed, if $O_{i}^{\prime }(\varphi ^{\prime })\equiv O_{i}(\varphi )$ we
obviously have 
\[
\langle O_{1}(\varphi (x_{1}))\cdots O_{n}(\varphi (x_{n}))\rangle =\langle
O_{1}^{\prime }(\varphi ^{\prime }(x_{1}))\cdots O_{n}^{\prime }(\varphi
^{\prime }(x_{n}))\rangle ^{\prime }, 
\]
where the primed average is calculated using the variables $\varphi ^{\prime
}$ and the unprimed average is calculated using the variables $\varphi $.
Certainly the information contained in these relations allows us to do
everything we need, with a suitable effort. On the other hand, working with
generating functionals we gain a compact formalism that makes most of that
effort for us. It would be very impractical to work without generating
functionals in non-Abelian gauge theories and gravity, for example, because
generating functionals give an easy control on local symmetries and their
properties under renormalization.

\section{Basic definitions and notation}

\setcounter{equation}{0}

Taking inspiration from the definitions introduced in ref. \cite{weinberg},
we call \textit{essential} a local composite field that is not a total
derivative and is not proportional to the field equations. When we say the a
composite field is ``proportional to the field equations'' we mean that it
is equal to the product of another composite field times $\delta S/\delta
\varphi $, or spacetime derivatives of $\delta S/\delta \varphi $. In all
other contexts when we say that an object is ``proportional to $X(\varphi )$%
'' we mean that it is equal to a constant times $X(\varphi )$.

Denote the essential composite fields with $\mathcal{O}^{\bar{I}}=\mathcal{O}%
^{\bar{I}}(\varphi ,\lambda )$. Call \textit{descendant} a composite field
that is a total derivative of an essential composite field. Define an
equivalence relation stating that two essential composite fields $\mathcal{O}%
^{\bar{I}}$ are equivalent if they differ by a descendant. Then, for each
equivalence class pick a representative $\mathcal{O}^{I}$ and couple it to a
source $L_{I}$. In the set of $\mathcal{O}^{I}$s we include the identity,
which is the ``composite field'' $1$, with source $L_{0}$, and the
elementary field $\varphi $ itself, with source $L_{1}$. It is convenient to
keep the source $J$ separate from the sources $L$, because $J$ identifies
the variables we are using to quantize the theory. Every perturbatively
local function $F(\varphi )$ of $\varphi $ can be expanded as a linear
combination of the form 
\begin{equation}
F(\varphi )=c_{I}(\lambda ,\partial )\mathcal{O}^{I}(\varphi ,\lambda
)+E(\varphi ).  \label{basicexpa}
\end{equation}
where $c_{I}(\lambda ,\partial )$ are operator-coefficients that may contain
derivatives acting on the objects that appear at their rights, and $%
E(\varphi )$ denotes terms proportional to the field equations. The sum over
repeated indices $I$ is understood. We say that $\{\mathcal{O}^{I}\}$ is a 
\textit{basis} of essential composite fields. We can choose a basis of
essential composite fields that may depend on the couplings, which is why we
use the notation $\mathcal{O}^{I}(\varphi ,\lambda )$.

For example, we can take the $\mathcal{O}^{I}$s with $I>1$ to be the
monomials constructed with $\varphi $ and its derivatives, discarding the
combinations that are equal to total derivatives or terms of type $E(\varphi
)$.

At the perturbative level, we write the action $S$ as the sum $S_{\text{free}%
}+S_{\text{int}}$ of its free- and interaction-parts, as usual. Then,
instead of identifying the terms of type $E(\varphi )$, we can equivalently
identify the terms $E_{\text{free}}(\varphi )$, proportional to the
free-field equations $\delta S_{\text{free}}/\delta \varphi $, and treat the
differences as linear combinations of other composite fields, to be
classified according to the same rules. This procedure is more convenient,
because it just requires to search for factors equal to (derivatives of) $%
(-\Box +m_{s}^{2})\varphi $ for scalars, $(\partial \!\!\!\slash+m_{f})\psi $
and $\bar{\psi}(-\overleftarrow{\partial \!\!\!\slash}+m_{f})$ for fermions,
and so on.

In some cases we may want to treat descendants and the terms of type $%
E(\varphi )$ or $E_{\text{free}}(\varphi )$ as independent composite fields
and add them to the basis $\{\mathcal{O}^{I}\}$. This redundancy may be
useful for several purposes. It is convenient to distinguish an \textit{%
essential} approach, where the basis $\{\mathcal{O}^{I}\}$ contains only
essential composite fields, and a \textit{redundant} approach, where the
basis $\{\mathcal{O}^{I}\}$ is unrestricted. We first work with the
redundant approach, because it is simpler, and then discuss the essential
approach in detail. Finally, we formulate a \textit{linear} approach, where
all sources $L_{I}$ renormalize linearly.

We denote the classical composite fields with $\mathcal{O}_{c}^{I}(\varphi
,\lambda )$. The classical basis $\{\mathcal{O}_{c}^{I}(\varphi ,\lambda )\}$
can be used to define the basis of bare essential composite fields $\{%
\mathcal{O}_{\mathrm{B}}^{I}(\varphi _{\mathrm{B}},\lambda _{\mathrm{B}})\}$%
, where the functions $\mathcal{O}_{\mathrm{B}}^{I}$ and $\mathcal{O}%
_{c}^{I} $ are just the same, but we call them with different names to
emphasize the fact that they have different arguments. Denote the bare
sources with $L_{I\mathrm{B}}$. We define the extended bare action $S_{L%
\mathrm{B}}$ as 
\begin{equation}
S_{L\mathrm{B}}(\varphi _{\mathrm{B}},\lambda _{\mathrm{B}},L_{\mathrm{B}%
})=S_{\mathrm{B}}(\varphi _{\mathrm{B}},\lambda _{\mathrm{B}})-\int L_{I%
\mathrm{B}}\mathcal{O}_{\mathrm{B}}^{I}(\varphi _{\mathrm{B}},\lambda _{%
\mathrm{B}}).  \label{SLB}
\end{equation}
The bare $Z$- and $W$-generating functionals are then 
\begin{equation}
Z_{\mathrm{B}}(J_{\mathrm{B}},\lambda _{\mathrm{B}},L_{\mathrm{B}})=\int [%
\mathrm{d}\varphi _{\mathrm{B}}]\hspace{0.02in}\exp \left( -S_{L\mathrm{B}%
}(\varphi _{\mathrm{B}},\lambda _{\mathrm{B}})+\int \varphi _{\mathrm{B}}J_{%
\mathrm{B}}\right) =\exp W_{\mathrm{B}}(J_{\mathrm{B}},\lambda _{\mathrm{B}%
},L_{\mathrm{B}}),  \label{ZLB}
\end{equation}
written in the conventional form.

Let us quickly review the renormalization in the presence of composite
fields. Relations (\ref{BRrel}) hold at $L_{I}=0$, so we need to concentrate
on the renormalization at $L_{I}\neq 0$. We define the basis $\{\mathcal{O}%
^{I}\}$ of composite fields 
\begin{equation}
\mathcal{O}^{I}(\varphi ,\lambda ,\mu )=\mathcal{O}_{\mathrm{B}}^{I}(\varphi
_{\mathrm{B}}(\varphi ,\lambda ,\mu ),\lambda _{\mathrm{B}}(\lambda ,\mu ))
\label{BRrel3}
\end{equation}
at the renormalized level. Note that these objects are not the renormalized
composite fields, but just the bare ones written using renormalized
variables and parameters. The renormalized composite fields will be
introduced later and denoted with $\mathcal{O}_{\mathrm{R}}^{I}$.

We start from the classical extended action 
\begin{equation}
S_{Lc}(\varphi ,\lambda ,\mu ,L)=S(\varphi ,\lambda ,\mu )-\int L_{I}%
\mathcal{O}^{I}(\varphi ,\lambda ,\mu ).  \label{SLB2}
\end{equation}
which is just the classical version of (\ref{SLB}). Here $L^{I}$ are the
renormalized sources for the composite fields. Then we write the functional
integral in the conventional form, using the action (\ref{SLB2}). Working
out the Feynman rules and calculating diagrams, we realize that physical
quantities are divergent. We calculate the divergent parts and subtract them
away \textit{modifying} the action as $S_{Lc}\rightarrow S_{L}=S_{Lc}+$
counterterms. We end up with a renormalized action $S_{L}(\varphi ,\lambda
,\mu ,L)$. The renormalized generating functionals $Z$ and $W$ are defined
as usual, once the action $S$ is replaced with $S_{L}$: 
\begin{equation}
Z(J,\lambda ,\mu ,L)=\int [\mathrm{d}\varphi ]\hspace{0.02in}\exp \left(
-S_{L}(\varphi ,\lambda ,\mu ,L)+\int \varphi J\right) =\exp W(J,\lambda
,\mu ,L).  \label{ZLR}
\end{equation}

Clearly, when we substitute $S_{Lc}$ with $S_{L}=S_{Lc}+$ counterterms we
are not making a change of variables, but just a replacement of actions. The
theory is written in the conventional form both before and after the
replacement, just because the term $\int J\varphi $ is replaced with the new
one by brute force.

We later realize that the bare and renormalized actions are related by
redefinitions of fields, parameters and sources, provided the classical
action contains sufficiently many independent parameters. Thus $S_{L}$ is
nothing but $S_{Lc}$, or $S_{\mathrm{B}}$, equipped with such redefinitions.
This only tells us that $S_{\mathrm{B}}$ and $S_{L}$ are related by such
redefinitions, which contain a change of variables for the fields, but not
that the functional integrals $Z_{\mathrm{B}}$ and $Z$ are also related by
those redefinitions. Indeed, we are not making any change of variables
inside the functional integral. Such a change of variables would affect the
term $\int J\varphi $ and convert the functional integral to some
unconventional form. Instead, we are jumping from the conventional
functional integral defined by the action $S_{\mathrm{B}}$ to the
conventional functional integral defined by the action $S_{L}$. At this
level, the BR map remains a replacement, not a true change of variables. In
the next sections we show how to upgrade it to a true change of variables.

\section{BR map in the redundant approach}

\setcounter{equation}{0}

In the redundant approach we work with a basis $\{\mathcal{O}^{I}\}$
containing all composite fields, including descendants (therefore also
derivatives of $\varphi $) and composite fields proportional to the field
equations. Then each perturbatively local function $F(\varphi )$ of $\varphi 
$ can be expanded as a linear combination 
\begin{equation}
F(\varphi )=c_{I}\mathcal{O}^{I}(\varphi ),  \label{basicexparedu}
\end{equation}
where the $c_{I}$s are constants. The renormalized action has the form 
\begin{equation}
S_{L}(\varphi ,\lambda ,\mu ,L)=S(\varphi ,\lambda ,\mu )-\int L_{I}\mathcal{%
O}^{I}(\varphi ,\lambda ,\mu )+\Delta S_{L}(\varphi ,\lambda ,\mu ,L),
\label{boroboro}
\end{equation}
where $\Delta S_{L}$ is a local functional that collects the counterterms
belonging to the composite-field sector. Expanding $\Delta S_{L}$ as shown
in (\ref{basicexparedu}) we can find local functions $f_{I}=L_{I}$ plus $%
\mathcal{O}(L)$-radiative corrections depending on the sources $L_{I}$ and
their derivatives, such that 
\begin{equation}
S_{L}(\varphi ,\lambda ,\mu ,L)=S(\varphi ,\lambda ,\mu )\,-\int
f_{I}(L,\lambda ,\mu )\mathcal{O}^{I}(\varphi ,\lambda ,\mu ).
\label{SLRredu}
\end{equation}

Then we see that the replacement 
\begin{equation}
\varphi _{\mathrm{B}}=\varphi _{\mathrm{B}}(\varphi ,\lambda ,\mu ),\qquad
\lambda _{\mathrm{B}}=\lambda _{\mathrm{B}}(\lambda ,\mu ),\qquad L_{I%
\mathrm{B}}=f_{I}(L,\lambda ,\mu ),\qquad \!\!\!\!\!\!\int J_{\mathrm{B}%
}\varphi _{\mathrm{B}}\leftrightarrow \int J\varphi ,  \label{BRreplaredu}
\end{equation}
turns the bare action (\ref{SLB}) into the renormalized one (\ref{SLR}) and
the bare generating functionals $Z_{\mathrm{B}}$ and $W_{\mathrm{B}}$ into
the renormalized ones $Z$ and $W$. We call (\ref{BRreplaredu}) the \textit{%
BR\ replacement} in the redundant approach.

Observe that the first of (\ref{BRreplaredu}) is just a \textit{classical}
change of variables. This means that it acts on the action as a change of
variables, but it is not meant as a change of variables in the functional
integral. The source $L_{0\mathrm{B}}$ reabsorbs field-independent
counterterms.

Now we see how to relate bare and renormalized quantities by a true change
of field variables. Use (\ref{BRrel}) to define the renormalized basis (\ref
{BRrel3}). Then,\ expand $\varphi _{\mathrm{B}}(\varphi ,\lambda ,\mu )$
using (\ref{basicexparedu}): 
\begin{equation}
\varphi _{\mathrm{B}}=\varphi +c_{I}\mathcal{O}^{I}(\varphi ,\lambda ,\mu ),
\label{BRrelexpa}
\end{equation}
where the constants $c_{I}$ can be treated perturbatively. Next, use this
relation to make a change of field variables in the bare functional integral
(\ref{ZLB}). Doing so, the term $\int J_{\mathrm{B}}\varphi _{\mathrm{B}}$
is turned into 
\[
\int J_{\mathrm{B}}\varphi +\int c_{I}J_{\mathrm{B}}\mathcal{O}^{I}(\varphi
), 
\]
so the functional integral is not written in the conventional form anymore.
We can convert it back to the conventional form if we define 
\begin{equation}
L_{I\mathrm{B}}=f_{I}(L)-c_{I}J,\qquad J_{\mathrm{B}}=J,  \label{blaredu}
\end{equation}
and apply these relations to (\ref{ZLB}), instead of using the last two
formulas of (\ref{BRreplaredu}). It is easy to see that we directly obtain 
\begin{equation}
Z_{\mathrm{B}}(J_{\mathrm{B}},\lambda _{\mathrm{B}},L_{\mathrm{B}%
})=Z(J,\lambda ,\mu ,L),\qquad W_{\mathrm{B}}(J_{\mathrm{B}},\lambda _{%
\mathrm{B}},L_{\mathrm{B}})=W(J,\lambda ,\mu ,L).  \label{coincide}
\end{equation}
Thus, in the redundant approach the change of variables converting bare
quantities into renormalized ones is 
\begin{equation}
\varphi _{\mathrm{B}}=\varphi _{\mathrm{B}}(\varphi ,\lambda ,\mu )=\varphi
+\sum_{I}c_{I}\mathcal{O}^{I},\qquad \lambda _{\mathrm{B}}=\lambda _{\mathrm{%
B}}(\lambda ,\mu ),\qquad L_{I\mathrm{B}}=f_{I}(L)-c_{I}J,\qquad J_{\mathrm{B%
}}=J.  \label{BRchangia}
\end{equation}
We call it the \textit{BR change of field variables}. Note that (\ref
{blaredu}) are linear in $J$, but not in $L$.

The replacement (\ref{BRreplaredu}) does not allow us to write the
equalities (\ref{coincide}). It just allows us to ``jump'' from the bare
generating functionals to the renormalized ones. For a variety of purposes
this ``jump'' is quite enough, yet it is not satisfactory if we want to
develop a field-covariant formulation of quantum field theory, where we need
operations such as (\ref{BRchangia}), which allow us to smoothly follow
every step of the transformation.

Before proceeding, let us make some observations about the functions $f^{I}$
in (\ref{SLRredu}). There exist constants $Z_{J}^{I}$ such that 
\[
f_{I}(L,\lambda ,\mu )=L_{J}(Z^{-1})_{I}^{J}+\mathcal{O}(L^{2}). 
\]
Indeed, at the linear level in $L$ we do not need to consider derivatives $%
\partial ^{p}L$, since integrating by parts these derivatives can be moved
inside the composite fields coupled to $L$. The terms $\mathcal{O}(L^{2})$
subtract the divergences that arise in correlation functions containing more
than one insertions of composite fields.

Taking the functional derivative of $Z$ and $W$ with respect to $L^{I}$ and
then setting $L=0$ we obtain convergent correlation functions containing one
insertion of $(Z^{-1})_{J}^{I}\mathcal{O}^{J}$ and arbitrarily many
insertions of elementary fields. Thus, $(Z^{-1})_{J}^{I}\mathcal{O}^{J}$ are
the renormalized composite fields $\mathcal{O}_{\mathrm{R}}^{I}$: 
\begin{equation}
\mathcal{O}_{\mathrm{R}}^{I}=(Z^{-1})_{J}^{I}\mathcal{O}^{J}(\varphi
,\lambda ,\mu ).  \label{oir}
\end{equation}
Using (\ref{BRrel3}) we get 
\[
\mathcal{O}_{\mathrm{B}}^{I}=Z_{J}^{I}\mathcal{O}_{\mathrm{R}}^{J}. 
\]
The quantities $Z_{J}^{I}$ are the renormalization constants of the
composite fields. It is convenient to organize the $\mathcal{O}^{I}$s in a
row such that operators of equal dimensions are close to one another and
operators of lower dimensions precede those of higher dimensions. In a
renormalizable theory, where only parameters of non-negative dimensions are
present, a composite field can mix only with composite fields of equal or
smaller dimensions, so the matrix $Z_{J}^{I}$ is block-lower triangular. If
the theory is non-renormalizable we can have two situations: if only
parameters of non-positive dimensions are present, then the matrix $%
Z_{J}^{I} $ is block-upper triangular; in the general case, where both
parameters of positive, null and negative dimensions are present, the matrix 
$Z_{J}^{I}$ has no particular restriction.

\section{BR replacement in the essential approach}

\setcounter{equation}{0}

We could be satisfied with the redundant approach just presented, however
for a variety of reasons that will be appreciated later we need to upgrade
it in several ways. It is often useful to restrict $\{\mathcal{O}^{I}\}$ to
be a basis of essential composite fields. Among the other things, this
allows us to better keep track of the counterterms proportional to the field
equations.

We prove that there exists a field redefinition $\varphi _{\mathrm{B}%
}=\varphi _{\mathrm{B}}(\varphi ,\lambda ,\mu ,L)$ such that the
renormalized action has the form 
\begin{equation}
S_{L}(\varphi ,\lambda ,\mu ,L)=S(\varphi ,\lambda ,\mu ,L)\,-\int
f_{I}(L,\lambda ,\mu )\mathcal{O}^{I}(\varphi ,\lambda ,\mu ),  \label{SLR}
\end{equation}
where $f_{I}=L_{I}$ plus radiative corrections are $\mathcal{O}(L)$-local
functions of the sources $L_{I}$ and their derivatives and 
\begin{equation}
S(\varphi ,\lambda ,\mu ,L)\,=S_{\mathrm{B}}(\varphi _{\mathrm{B}}(\varphi
,\lambda ,\mu ,L),\lambda _{\mathrm{B}}(\lambda ,\mu )).  \label{slr2}
\end{equation}
Note that the functions $f_{I}$ are not the same as in (\ref{SLRredu}),
although we use the same notation, for simplicity. At $L_{I}\neq 0$ the
field replacement is the first of (\ref{BRrel}) plus $\mathcal{O}(L)$%
-radiative corrections to be determined. Thus, $S_{L}$ is equal to the
classical extended action $S_{Lc}$ of (\ref{SLB2}) plus radiative
corrections.

We already know that at $L_{I}=0$ renormalization is given by (\ref{BRrel}).
At $L_{I}\neq 0$ we proceed iteratively in the loop expansion, as usual. At
a given order, we treat the action as an expansion in powers of $L$.
Assuming that divergences are subtracted away up to $n-1$ loops by a
renormalized action of the form (\ref{SLR}), we study the $n$-loop $\mathcal{%
O}(L)$-counterterms. It is convenient to expand them as shown in (\ref
{basicexpa}). We can distinguish counterterms proportional to the field
equations and counterterms proportional to the essential fields and their
derivatives. Integrating by parts we can write the $n$-loop $\mathcal{O}(L)$%
-divergences as the sum 
\begin{equation}
\int \Delta _{I}(L)\mathcal{O}^{I}(\varphi ,\lambda ,\mu )+\int U(\varphi ,L)%
\frac{\delta S}{\delta \varphi },  \label{counte}
\end{equation}
where $U(\varphi ,L)$ is an $n$-loop $\mathcal{O}(L)$-local function of $%
\varphi $, $L$ and their derivatives and the $\Delta _{I}(L)$s are $n$-loop $%
\mathcal{O}(L)$-local functions of the sources $L$ and their derivatives.

We subtract the first sum of (\ref{counte}) renormalizing the sources $L_{I}$
by means of the replacements 
\begin{equation}
L_{I}\rightarrow L_{I}+\Delta _{I}(L).  \label{LD}
\end{equation}
However, when we perform this replacement inside (\ref{SLR}) we not only
subtract the first term of (\ref{counte}), but generate also higher-loop
divergent terms. They can be ignored at this stage. They will be dealt with
at the subsequent steps of the subtraction algorithm. As far as the $n$-loop
counterterms are concerned, inserting (\ref{LD}) inside (\ref{SLR}) or (\ref
{SLB2}) is the same thing.

Consider now the counterterms proportional to the field equations. We
subtract them redefining the fields $\varphi $ as $\varphi =\varphi ^{\prime
}-U(\varphi ^{\prime },L)$ inside $S_{L}$ and replacing the term $\int
J\varphi $ with $\int J\varphi ^{\prime }$. Later we drop the prime on $%
\varphi ^{\prime }$ and rename $\varphi ^{\prime }$ as $\varphi $. Again,
these operations are just replacements, not changes of variables inside the
functional integral.

As above, when we perform the replacement $\varphi \rightarrow \varphi -U$
inside the action we not only subtract the last term of (\ref{counte}), but
generate also divergent terms proportional to higher powers of $U$, to be
dealt with at the subsequent steps of the subtraction algorithm. However,
the replacement $\varphi \rightarrow \varphi -U$ also affects the essential
fields. Neglecting higher-order terms, we can focus on the $n$-loop
contributions 
\begin{equation}
-\int \mathrm{d}x\ L_{I}(x)\frac{\delta \mathcal{O}^{I}(x)}{\delta \varphi
(y)}U(y)\ \mathrm{d}y,  \label{divop}
\end{equation}
which are now $\mathcal{O}(L^{2})$. These terms can be treated like the
divergences (\ref{counte}). Again, we start separating the part proportional
to the field equations from the part proportional to essential composite
fields and descendants, and repeat the procedure. We get an $\Delta _{I}(L)$
and a $U(\varphi ,L)$ of order $L^{2}$. With another $L_{I}$-replacement
like (\ref{LD}) and a field replacement, we remain with $n$-loop divergent
terms (\ref{divop}) proportional to $\mathcal{O}(L^{3})$. Proceeding
indefinitely in this way, we can get rid of all of them.

Then we repeat the entire procedure for the $(n+1)$-loop $\mathcal{O}(L)$%
-counterterms, and so on. This iteration proves our statement. The
replacements (\ref{LD}) build the functions $f_{I}$. The field replacements
build the functions $\varphi _{\mathrm{B}}(\varphi ,\lambda ,\mu ,L)$.

Summarizing, there exist \textit{BR replacements} 
\begin{equation}
\varphi _{\mathrm{B}}=\varphi _{\mathrm{B}}(\varphi ,\lambda ,\mu ,L),\qquad
\!\!\!\!\!\!\lambda _{\mathrm{B}}=\lambda _{\mathrm{B}}(\lambda ,\mu
),\qquad L_{I\mathrm{B}}=f_{I}(L,\lambda ,\mu ),\qquad \int J_{\mathrm{B}%
}\varphi _{\mathrm{B}}\leftrightarrow \int J\varphi ,  \label{BRrepla}
\end{equation}
that turn the bare action (\ref{SLB}) into the renormalized one (\ref{SLR})
and the bare generating functionals $Z_{\mathrm{B}}$ and $W_{\mathrm{B}}$
into the renormalized ones $Z$ and $W$.

We stress again that formulas (\ref{SLB}) and (\ref{SLR}), as well as (\ref
{BRreplaredu}) and (\ref{BRrepla}) are very general, and hold even when the
relation between bare and renormalized fields is non-polynomial, and the
theory is non-renormalizable.

Observe that shifting some $L$s by constants is equivalent to modify the
action $S(\varphi ,\lambda ,\mu )$. Using this trick we can turn on
non-renormalizable vertices, for example. Renormalizable interactions can
also be described this way, starting from a free-field theory. This is not
surprising, because studying composite fields in a free-field theory we get
enough information to reconstruct every perturbatively interacting theory,
renormalizable or not. Basically, what we do is to replace the classical
action $S_{c}$ with the free-field action $S_{\text{free}}$ and shift the
sources $L$ by constants, such that the extended action $S_{Lc}$ of (\ref
{SLB2}) at $L=0$ gives back $S_{c}$. Similar operations can be made in the
bare action (\ref{SLB}) and the renormalized action (\ref{SLR}). Using this
trick we can also retrieve (\ref{BRrel}) from (\ref{BRrepla}). For example,
the massive $\varphi ^{4}$-theory can be described working with the free
action $S_{\text{free}}(\varphi )$ and setting all sources $L$ equal to zero
except for the sources $L_{4}$ and $L_{2}$ of $\varphi ^{4}/4!$ and $\varphi
^{2}/2!$, which are set equal to the constants $-\lambda $ and $-m^{2}$,
respectively. Then the relations $L_{I\mathrm{B}}=f_{I}(L,\lambda ,\mu )$ of
(\ref{BRrepla}) with $I=4,2$, become $\lambda _{\mathrm{B}}=\lambda _{%
\mathrm{B}}(\lambda ,\mu ),$ and $m_{\mathrm{B}}^{2}=m^{2}Z_{m}(\lambda ,\mu
)$, while $\varphi _{\mathrm{B}}=\varphi _{\mathrm{B}}(\varphi ,\lambda ,\mu
,L)$ becomes the first of (\ref{BRrel}), which in this case is nothing but $%
\varphi _{\mathrm{B}}=Z_{\varphi }^{1/2}\varphi $.

\section{BR change of variables in the essential approach}

\setcounter{equation}{0}

Now we study the relation between bare and renormalized quantities in the
essential approach as a change of variables in the functional integral. We
show that there exist local functions 
\begin{equation}
J_{\mathrm{B}}=J,\qquad L_{I\mathrm{B}}=L_{I\mathrm{B}}(J,\lambda ,\mu ,L),
\label{cambio}
\end{equation}
relating bare and renormalized sources, that make the bare and renormalized $%
Z$- and $W$-functionals coincide. Moreover, we give an algorithm to derive
the functions (\ref{cambio}) explicitly.

Using the first of (\ref{BRrepla}) write the action $S(\varphi ,\lambda ,\mu
,L)$ of (\ref{slr2}). Next, expand the relation between $\varphi _{\mathrm{B}%
}$ and $\varphi $ as shown in (\ref{basicexpa}). We have 
\begin{equation}
\varphi _{\mathrm{B}}(\varphi ,\lambda ,\mu ,L)=\varphi +c_{I}(\lambda ,\mu
,L,\partial )\mathcal{O}^{I}(\varphi ,\lambda ,\mu )+E(\varphi ),
\label{fibb}
\end{equation}
which is equal to $\varphi $ plus perturbative corrections. Here $E(\varphi
) $ is meant to be proportional to the field equations of $S(\varphi
,\lambda ,\mu ,L)$. The coefficients $c_{I}$ are local functions of the
sources $L$ and their derivatives, and can contain derivative operators
acting on their right-hand sides. Similarly, inserting (\ref{fibb}) in the
bare composite fields, using (\ref{BRrel3}) and expanding again, we can
write 
\[
\mathcal{O}_{\mathrm{B}}^{I}(\varphi _{\mathrm{B}},\lambda _{\mathrm{B}%
})=d_{J}^{I}(\lambda ,\mu ,L,\partial )\left( \mathcal{O}^{J}(\varphi
,\lambda ,\mu )+E^{J}(\varphi )\right) \text{,} 
\]
where $d_{J}^{I}=\delta _{J}^{I}+\mathcal{O}(L)$. Next, use (\ref{fibb}) to
make a change of variables in the functional integral $Z_{\mathrm{B}}$ of (%
\ref{ZLB}). At the same time, write 
\begin{equation}
L_{I\mathrm{B}}=(d^{-1}(-\partial ))_{I}^{J}\left( f_{J}-c_{J}(-\partial
)J+\Delta f_{J}\right) ,\qquad J_{\mathrm{B}}=J,  \label{bla}
\end{equation}
where $\Delta f_{I}$ are yet unknown $\mathcal{O}(\tilde{L}^{2})$-local
functions of the sources $\tilde{L}=\{L,J\}$. We recall that $f_{I}$ are $%
\mathcal{O}(L)$-local functions of $L$, $\lambda $, and $\mu $.

As usual, using the dimensional-regularization technique, which we assume
here, and treating the change of variables perturbatively, the functional
integration measure is invariant. Everything works as in the replacement of
the previous section, but for the term $\int \varphi _{\mathrm{B}}J_{\mathrm{%
B}}$ and the corrections proportional to $\Delta f_{I}$. The exponent $-S_{L%
\mathrm{B}}+\int \varphi _{\mathrm{B}}J_{\mathrm{B}}$ turns into minus (\ref
{SLR}) plus 
\begin{equation}
\int J\varphi +\int \Delta f_{I}\mathcal{O}^{I}-\int U(\varphi ,J,L)\frac{%
\delta S}{\delta \varphi },  \label{middle}
\end{equation}
for some local function $U(\varphi ,J,L)=\mathcal{O}(\tilde{L})$. The first
term is the one that must be there, while the rest must be canceled out. The
second integral in (\ref{middle}) is $\mathcal{O}(\tilde{L}^{2})$ by
assumption, but yet unknown. The last term of (\ref{middle}) collects all
contributions proportional to the field equations. This object can be
manipulated with the procedure described after formula (\ref{counte}). The
only difference is that now the field redefinitions are true change of
variables inside the functional integral. Thus, make the change of variables 
$\varphi =\varphi ^{\prime }-U(\varphi ^{\prime },J,L)$ and then drop the
prime on $\varphi ^{\prime }$ to rename $\varphi ^{\prime }$ as $\varphi $.
Expanding in powers of $U$ we cancel the last term of (\ref{middle}), but
generate new terms, which, however, are all $\mathcal{O}(\tilde{L}^{2})$.
These terms include the object $-\int JU$ originating from $\int J\varphi $.
Then we expand such terms using (\ref{basicexpa}), and cancel essential
composite fields and descendants fixing the $\mathcal{O}(\tilde{L}^{2})$%
-contributions to $\Delta f_{I}\hspace{0.01in}$. After this, the last term
of formula (\ref{middle}) is replaced by an object of the same form, but one
order of $\tilde{L}$ higher than before. Then we can repeat the procedure.
Iterating the procedure indefinitely we obtain (\ref{coincide}).

Observe that in the end the redefinitions (\ref{cambio}) can contain
arbitrarily large powers of $L$ and $J$. Moreover, combining the first of (%
\ref{BRrepla}) with the further changes of variables of type $\varphi
=\varphi ^{\prime }-U$, the final change of variables that relates bare and
renormalized fields is $J$-dependent, so instead of the first of (\ref
{BRrepla}) we have $\varphi _{\mathrm{B}}=\varphi _{\mathrm{B}}(\varphi
,\lambda ,\mu ,L,J)$.

Recapitulating, the BR change of variables in the essential approach has the
form 
\begin{equation}
\varphi _{\mathrm{B}}=\varphi _{\mathrm{B}}(\varphi ,\lambda ,\mu
,L,J),\qquad \lambda _{\mathrm{B}}=\lambda _{\mathrm{B}}(\lambda ,\mu
),\qquad L_{I\mathrm{B}}=L_{I\mathrm{B}}(J,\lambda ,\mu ,L),\qquad J_{%
\mathrm{B}}=J,  \label{BRchange}
\end{equation}
where $\varphi _{\mathrm{B}}=\varphi +$ radiative corrections, $L_{I\mathrm{B%
}}=L_{I}+$ $\mathcal{O}(\tilde{L})$-radiative corrections, and so on.
Formula (\ref{BRchange}) allows us to describe the relation between bare and
renormalized quantities as a true change of variables in the functional
integral, instead of using the replacement (\ref{BRrepla}).

Note that the source $J$ never renormalizes. This happens because its
renormalization is moved to the renormalization of the source $L_{1}$
coupled with the elementary field $\varphi $. For example, in the massive $%
\varphi ^{4}$ theory we get 
\begin{eqnarray}
\varphi _{\mathrm{B}} &=&\varphi Z_{\varphi }^{1/2}(\lambda ,\mu ),\qquad
\lambda _{\mathrm{B}}=\lambda \mu ^{\varepsilon }Z_{\lambda }(\lambda ,\mu
),\qquad m_{\mathrm{B}}^{2}=m^{2}Z_{m}(\lambda ,\mu ),  \nonumber \\
L_{1\mathrm{B}} &=&L_{1}Z_{\varphi }^{-1/2}+J(Z_{\varphi }^{-1/2}-1),\qquad
J_{\mathrm{B}}=J,  \label{bubub}
\end{eqnarray}
where $Z_{\lambda }$ and $Z_{m}$ are the renormalization constants of the
coupling and the squared mass, respectively. A map similar to (\ref{bubub})
holds in every multiplicatively renormalizable theory. The functional
integral depends only on $J+L_{1}$, at this level, however in other
approaches (see next section) and other applications $J$ and $L_{1}$ play
different roles, which is why we prefer to keep them distinct.

\section{BR map in the linear approach}

\label{linearapproach}\setcounter{equation}{0}

The essential approach takes advantage of the most general field- and
source-redefinitions and makes us appreciate the role played by
higher-powers of $J$ in (\ref{cambio}), as well as the roles played by $J$-
and $L$-dependences in the relation $\varphi _{\mathrm{B}}\leftrightarrow
\varphi $. However, BR replacements and BR changes of variables are much
simpler if we adopt the redundant approach. There the source redefinitions
are linear in $J$, although not in $L$.

The reader may be worried that the redundant approach does not isolate the
counterterms that can be removed by means of field redefinitions. This is
true, but only in the source sector, because the source-independent sector
is taken care by (\ref{BRrel}). Nothing prevents us from removing the
divergences proportional to the field equations by means of field
redefinitions in the source-independent sector, even if we use the redundant
approach. For most purposes, this is enough. Indeed, isolating terms
proportional to the field equations in the source-independent sector is
necessary to identify key properties of the theory, such as its finiteness,
or its renormalizability with a finite number of independent parameters.
Instead, isolating terms proportional to the field equations in the source
sector is more a matter of aesthetics.

These arguments lead us to conclude that the redundant approach is more
convenient than the essential one. We can make a step forward and define a
third approach, which is even more convenient for several purposes. We call
it the \textit{linear approach}, because all source redefinitions are linear
in $L$ as well as in $J$.

The bare action is not written in the form (\ref{SLB}), rather in the new
form 
\begin{equation}
S_{L\mathrm{B}}(\varphi _{\mathrm{B}},\lambda _{\mathrm{B}},\tau _{\mathrm{B}%
},L_{\mathrm{B}})=S_{\mathrm{B}}(\varphi _{\mathrm{B}},\lambda _{\mathrm{B}%
})-\int (L_{I\mathrm{B}}+\tau _{vI\mathrm{B}}\mathcal{N}_{\mathrm{B}}^{v}(L_{%
\mathrm{B}},\lambda _{\mathrm{B}}))\mathcal{O}_{\mathrm{B}}^{I}(\varphi _{%
\mathrm{B}},\lambda _{\mathrm{B}}),  \label{SLBlinear}
\end{equation}
where $\mathcal{N}_{\mathrm{B}}^{v}(L_{\mathrm{B}},\lambda _{\mathrm{B}})=%
\mathcal{O}(L_{\mathrm{B}}^{2})$ is a basis of independent local monomials
that can be constructed with the sources $L_{\mathrm{B}}$ and their
derivatives, and are at least quadratic in $L_{\mathrm{B}}$. Each such
monomial is multiplied by a new, independent coefficient $\tau _{vI\mathrm{B}%
}$. The sum over repeated indices $v$ is understood. Note that in the linear
approach the functional integral depends on $J$ and $L_{1}$ separately. The
bare $Z$- and $W$-functionals are given by (\ref{ZLB}), with the extended
action (\ref{SLBlinear}).

The classical action is (\ref{SLBlinear}) once the subscripts B are dropped, 
\begin{equation}
S_{cL}(\varphi ,L)=S_{c}(\varphi )-\int L_{I}\mathcal{O}_{c}^{I}(\varphi
)-\int \tau _{vI}\mathcal{N}^{v}(L)\mathcal{O}_{c}^{I}(\varphi ).
\label{bibop}
\end{equation}
The renormalized action $S_{L}(\varphi ,\lambda ,\tau ,L)$ is derived below,
check formulas (\ref{esln}) and (\ref{esln4}).

Before proceeding, let us explain how the perturbative expansion must be
organized. We want to be sure that radiative corrections are of higher order
with respect to the classical terms. This fact is obvious when composite
fields are switched off, less obvious when they are present. We describe the
behavior of each quantity referring it to some parameter $\delta \ll 1$. Let
us state that the coupling $\lambda _{n_{l}}$ multiplying a vertex with $%
n_{l}$ $\varphi $-legs is of order $\delta ^{n_{l}-2}$. Then when composite
fields are switched off each loop carries an additional factor $\delta ^{2}$%
. Consider an $\ell $-loop diagram with $E$ external legs, $I$ internal legs
and $v_{l}$ vertices of type $l$. Counting legs and using the identity $\ell
=I-V+1$, we have $\sum_{l}n_{l}v_{l}=E+2I=E+2(\ell +V-1)$, so the diagram is
multiplied by an expression of order 
\begin{equation}
\delta ^{\sum_{l}(n_{l}-2)v_{l}}=\delta ^{2\ell }\delta ^{E-2}.  \label{ax}
\end{equation}
Besides the expected tree-level factor $\delta ^{E-2}$, associated with the $%
E$ external $\varphi $-legs, we get a $\delta ^{2}$ for each loop, as
claimed.

Now, assume that the composite fields $\mathcal{O}^{I}(\varphi ,\lambda )$
are homogeneous in $\delta $, namely 
\[
\mathcal{O}^{I}(\varphi \delta ^{-1},\lambda _{l}\delta ^{n_{l}-2})=\delta
^{-n_{I}}\mathcal{O}^{I}(\varphi ,\lambda ) 
\]
for some $n_{I}$. Observe that the vertices $L_{I}\mathcal{O}^{I}$ are not
multiplied by any coupling. Actually, the sources $L$ replace the couplings
in this case, so we must assume $L_{I}=\mathcal{O}(\delta ^{n_{I}-2})$.
Next, consider the parameters $\tau $ and observe that some contributions to
their renormalization can be $\mathcal{O}(\delta ^{0})$, because the
vertices $L_{I}\mathcal{O}^{I}$ allow us to construct diagrams with no
couplings $\lambda $ and at least two external $L$-legs. Thus, the
parameters $\tau $ may carry negative orders of $\delta $. A consistent
assignment is $\tau _{vI}=\mathcal{O}(\delta ^{n_{I}-n_{v}-2})$, where $%
n_{v} $ is the $\delta $-degree of the monomial $\mathcal{N}^{v}(L)$,
because then the product $\tau _{vI}\mathcal{N}^{v}$ is $\mathcal{O}(\delta
^{n_{I}-2})$, like $L_{I}$.

Summarizing, the $\delta $-expansion is properly organized assuming 
\begin{equation}
\lambda _{n_{l}}=\mathcal{O}(\delta ^{n_{l}-2}),\qquad L_{I}=\mathcal{O}%
(\delta ^{n_{I}-2}),\qquad \tau _{vI}=\mathcal{O}(\delta ^{n_{I}-n_{v}-2}),
\label{assigne0}
\end{equation}
while $J$ is $\mathcal{O}(\delta ^{-1})$. A quick way to derive the correct
assignments is to observe that if the fields $\varphi $ are imagined to be $%
\mathcal{O}(\delta ^{-1})$, then all terms of the classical action are $%
\mathcal{O}(\delta ^{-2})$. In particular, if we make the substitutions 
\begin{equation}
\varphi \rightarrow \varphi \delta ^{-1},\qquad \lambda _{n_{l}}\rightarrow
\lambda _{n_{l}}\delta ^{n_{l}-2},\qquad L_{I}\rightarrow L_{I}\delta
^{n_{I}-2},\qquad \tau _{vI}\rightarrow \tau _{vI}\delta ^{n_{I}-n_{v}-2},
\label{substituti}
\end{equation}
then the action transforms as 
\begin{equation}
S_{L}(\varphi ,\lambda ,\tau ,L)\rightarrow \frac{1}{\delta ^{2}}\left(
S_{cL}+\sum_{\ell \geqslant 1}\delta ^{2\ell }S_{\ell L}\right) ,
\label{substitutii}
\end{equation}
where the $\ell $-loop contributions $S_{\ell L}$ are $\delta $-independent.

Sometimes it may be useful to consider some couplings of orders higher than
those assigned in (\ref{assigne0}). This is allowed, depending on the
specific features of the theory, as long as the radiative corrections to
those couplings are of even higher orders.

Equipped with the more involved structures (\ref{SLBlinear}) and (\ref{bibop}%
), renormalization is now much simpler. We can define a redundant linear
approach and an essential linear approach. We begin with the redundant one.

The action (\ref{bibop}) contains enough independent parameters to
renormalize all $\mathcal{O}(L^{2})$-divergences relating $\tau _{\mathrm{B}%
} $ and $\tau $. Doing so it is sufficient to renormalize the sources $L_{I%
\mathrm{B}}$ linearly, using the renormalization constants $Z_{IJ}$ already
met: 
\[
L_{I\mathrm{B}}=L_{J}(Z^{-1})_{I}^{J}. 
\]
Thus, in the redundant linear approach the BR replacement that turns (\ref
{ZLB}) into (\ref{ZLR}) reads 
\begin{eqnarray}
\varphi _{\mathrm{B}} &=&\varphi _{\mathrm{B}}(\varphi ,\lambda ,\mu
),\qquad \lambda _{\mathrm{B}}=\lambda _{\mathrm{B}}(\lambda ,\mu ),\qquad
\int J_{\mathrm{B}}\varphi _{\mathrm{B}}\leftrightarrow \int J\varphi , 
\nonumber \\
L_{I\mathrm{B}} &=&L_{J}(Z^{-1})_{I}^{J},\qquad \tau _{vI\mathrm{B}}=\hat{%
\tau}_{vJ}(\tau ,\lambda ,\mu )(Z^{-1})_{I}^{J},  \label{BRreplareduline}
\end{eqnarray}
instead of (\ref{BRreplaredu}), where $\hat{\tau}=\tau $ plus radiative
corrections. The renormalized extended action is 
\begin{equation}
S_{L}(\varphi ,\lambda ,\mu ,L)=S(\varphi ,\lambda ,\mu )-\int \left( L_{I}+%
\hat{\tau}_{vI}\mathcal{N}^{v}(L,\lambda ,\mu )\right) \mathcal{O}_{\mathrm{R%
}}^{I}(\varphi ,\lambda ,\mu ),  \label{esln}
\end{equation}
where $\mathcal{O}_{\mathrm{R}}^{I}$ are the renormalized composite fields (%
\ref{oir}) and we have defined the basis of renormalized $\mathcal{N}^{v}$s
as $\mathcal{N}^{v}(L,\lambda ,\mu )=\mathcal{N}_{\mathrm{B}}^{v}(L_{\mathrm{%
B}},\lambda _{\mathrm{B}})$. It is easy to check that (\ref{BRreplareduline}%
) is consistent with the perturbative expansion governed by (\ref{assigne0}).

To study the BR map as a change of field variables we first make the change
of variables (\ref{BRrelexpa}) inside the functional integral. The constants 
$c_{I}$ must be assumed to be of order $\delta ^{n_{I}+1}$. A factor $\delta
^{n_{I}-1}$ is the tree-level assignment that makes $\varphi $ and $c_{I}%
\mathcal{O}^{I}$ of the same order. An extra factor $\delta ^{2}$ comes from
the fact that the $c_{I}$s are at least one loop. (In the next sections we
also consider classical changes of variables, where the $c_{I}$s become of
order $\delta ^{n_{I}-1}$.) We also make the substitutions 
\begin{equation}
L_{I\mathrm{B}}=L_{J}(Z^{-1})_{I}^{J}-c_{I}J+L_{J}h_{I}^{J}-c_{J}J\Delta
c_{I}^{J},\qquad J_{\mathrm{B}}=J,\qquad \tau _{vI\mathrm{B}}=(\hat{\tau}%
_{vJ}+\Delta \hat{\tau}_{vJ})(Z^{-1})_{I}^{J},  \label{linearepla}
\end{equation}
where $h_{I}^{J}$, $\Delta c_{I}^{J}$ and $\Delta \hat{\tau}_{v}$ are
unknown constants. They have to be determined as expansions in powers of $c$
starting with $\mathcal{O}(c)$, so they can be treated perturbatively. The
exponent of the $Z$-integrand contains the functions $\mathcal{N}^{v}$,
which generate other $J$-dependent terms after the replacements (\ref
{linearepla}). Such terms have the structure 
\begin{equation}
\int Jc_{I}U^{I}(\varphi ,cJ,L,h,\Delta c,\Delta \hat{\tau}),  \label{astrot}
\end{equation}
where $U^{I}$ are $\mathcal{O}(\tilde{L})$-local functions of $\tilde{L}%
=\{L,J\}$. The exponent of the $Z$-integrand can be written as minus (\ref
{esln}) plus $\int J\varphi $ plus 
\begin{equation}
\int \left( L_{J}h_{I}^{J}-c_{J}J\Delta c_{I}^{J}\right) \mathcal{O}%
^{I}+\int \left( \hat{\tau}_{vI}(C_{w}^{v}-\delta _{w}^{v})+\Delta \hat{\tau}%
_{vI}C_{w}^{v}\right) \mathcal{N}^{w}(L,\lambda ,\mu )\mathcal{O}_{\mathrm{R}%
}^{I}+\int Jc_{I}U^{I},  \label{erto}
\end{equation}
where we have expanded 
\[
\mathcal{N}_{\mathrm{B}}^{v}(L(Z^{-1}+h),\lambda _{\mathrm{B}})=C_{w}^{v}%
\mathcal{N}^{w}(L,\lambda ,\mu ), 
\]
and $C_{w}^{v}=\delta _{w}^{v}+\mathcal{O}(h)$ are constants.

The terms (\ref{erto}) are those we must get rid of in order to obtain the
renormalized generating functionals (\ref{ZLR}) and prove relations (\ref
{coincide}). Now we show that we can achieve this goal choosing $h$, $\Delta
c_{I}$ and $\Delta \hat{\tau}_{vI}$ appropriately.

Make the further change of variables $\varphi \rightarrow \varphi
-c_{I}U^{I} $. Expanding in the basis $\{\mathcal{O}^{I}\}$ and $\{\mathcal{N%
}^{v}\}$, the action $S(\varphi ,\lambda ,\mu )$ transforms as 
\begin{equation}
S(\varphi ,\lambda ,\mu )\rightarrow S(\varphi ,\lambda ,\mu )+\int \left(
L_{J}\bar{h}_{\hspace{0.01in}I}^{J}-c_{I}J\bar{\Delta}c_{I}^{J}\right) 
\mathcal{O}^{I}+\int \bar{\Delta}\hat{\tau}_{vI}\mathcal{N}^{v}(L,\lambda
,\mu )\mathcal{O}_{\mathrm{R}}^{I}-c^{2}\int J\tilde{U}(\varphi ,cJ,L),
\label{erto4}
\end{equation}
where the last term is written in compact form (indices being understood), $%
\bar{\Delta}c$, $\bar{h}$ and $\bar{\Delta}\hat{\tau}$ are $\mathcal{O}(c)$%
-constants and $\tilde{U}=\mathcal{O}(\tilde{L})$ are local functions. All
these objects are perturbative expansions in powers of $c$, whose
coefficients may also depend on $h$, $\Delta c_{I}$ and $\Delta \hat{\tau}%
_{vI}$. An expansion similar to (\ref{erto4}) can be written for the
transformed $\mathcal{O}^{I}(\varphi ,\lambda ,\mu )$s, but since these
objects are always multiplied by $\mathcal{O}(\tilde{L})$, they affect the
exponent only adding terms like the last two of (\ref{erto4}). The
corrections originating from (\ref{astrot}) only give terms like the last of
(\ref{erto4}). Finally, the exponent of the $Z$-integrand is minus (\ref
{esln}) plus $\int J\varphi $ plus 
\begin{eqnarray}
&&\int \left( L_{J}(h_{I}^{J}-\bar{h}_{\hspace{0.01in}I}^{J})-c_{I}J(\Delta
c_{I}^{J}-\bar{\Delta}c_{I}^{J})\right) \mathcal{O}^{I}+\int \left( \hat{\tau%
}_{vI}(C_{w}^{v}-\delta _{w}^{v})+\Delta \hat{\tau}_{vI}C_{w}^{v}-\bar{\Delta%
}\hat{\tau}_{vI}^{\prime }\right) \mathcal{N}^{w}(L,\lambda ,\mu )\mathcal{O}%
_{\mathrm{R}}^{I}  \nonumber \\
&&\qquad \qquad +c^{2}\int JU^{\prime }(\varphi ,cJ,L),  \label{erto5}
\end{eqnarray}
with $U^{\prime }=\mathcal{O}(\tilde{L})$. The first line of (\ref{erto5})
is canceled choosing the unknowns $h_{I}^{J}$, $\Delta c_{I}^{J}$ and $%
\Delta \hat{\tau}_{v}$, which can be done solving their equations
recursively in powers of $c$. Actually, at this stage we can truncate the
solutions at $\mathcal{O}(c)$, because higher orders must be modified
anyway, to cancel the second line of (\ref{erto5}). Such terms are like the
term $\int JcU$ of (\ref{astrot}), but one order higher in $c$. Then we
repeat the procedure, starting from the change of variables $\varphi
\rightarrow \varphi -c^{2}U^{\prime }$, and determine higher-order
corrections to the constants $h_{I}^{J}$, $\Delta c_{I}$ and $\Delta \hat{%
\tau}_{vI}$. Proceeding indefinitely like this, we get (\ref{ZLR}) and
relations (\ref{coincide}).

Summarizing, the BR change of variables in the redundant linear approach has
the form 
\begin{eqnarray}
\varphi _{\mathrm{B}} &=&\varphi _{\mathrm{B}}(\varphi ,\lambda ,\mu
,J,L),\qquad \lambda _{\mathrm{B}}=\lambda _{\mathrm{B}}(\lambda ,\mu
),\qquad \!\!\!\!\!\!\tau _{\mathrm{B}}=\tilde{\tau}(\tau ,\lambda ,\mu ), 
\nonumber \\
L_{I\mathrm{B}} &=&(L_{J}-\tilde{c}_{J}J)(\tilde{Z}^{-1})_{I}^{J},\qquad J_{%
\mathrm{B}}=J,  \label{blareduline}
\end{eqnarray}
where 
\begin{eqnarray*}
&&\varphi _{\mathrm{B}}(\varphi ,\lambda ,\mu ,J,L)=\varphi +c_{I}\mathcal{O}%
^{I}+c_{I}\hspace{0.01in}\hat{U}^{I}(\varphi ,cJ,L),\qquad \hat{U}^{I}=%
\mathcal{O}(\tilde{L}), \\
&(\tilde{Z}^{-1})_{I}^{J}=&(Z^{-1})_{I}^{J}+\mathcal{O}(c),\qquad \tilde{c}%
_{I}=c_{I}+\mathcal{O}(c^{2}),\qquad \tilde{\tau}_{vI}=\hat{\tau}%
_{vJ}(Z^{-1})_{I}^{J}+\mathcal{O}(c).
\end{eqnarray*}
The $Z$- and $W$-functionals behave as scalars. The crucial property of the
linear approach is that the second line of (\ref{blareduline}) is linear in
both $J$ and $L$.

\bigskip

Now we study the essential linear approach. Here we need to eliminate
descendants and terms proportional to the field equations. Descendants are
taken care of converting the constants $c$, $Z$, $\tilde{c}$ and $\tilde{Z}$
into derivative-operators. We do not do this explicitly, because it is
straightforward. The linearity of $L_{I\mathrm{B}}$ in both $J$ and $L$ is
preserved in this extended sense. This trick can also help us eliminate the
terms $E(\varphi )$ proportional to the field equations, if we identify such
terms as $E_{\text{free}}(\varphi )$ plus perturbative corrections. Indeed, $%
E_{\text{free}}(\varphi )$ are just descendants plus mass terms. If we use
this trick, the redundant and essential linear approaches practically
coincide.

Instead, if we want to eliminate the terms $E(\varphi )$ belonging to the
source sector by means of further field redefinitions, the procedure we must
apply is the same as the one we used to eliminate the last term of (\ref
{middle}), expanding in powers of $L$ or $\tilde{L}$. We briefly describe it
here.

Let us begin with the BR\ replacement. The terms proportional to the field
equations we want to reabsorb are $\mathcal{O}(L)$. They can be canceled by
means of $\mathcal{O}(L)$-corrections to the field replacement, which,
however, generate also other $\mathcal{O}(L^{2})$-terms. These can be
expanded in the basis $\mathcal{O}^{I}$, and their coefficients can be
expanded in the basis $\mathcal{N}^{v}$, and canceled redefining the
constants $\hat{\tau}(\tau ,\lambda ,\mu )$, up to $\mathcal{O}(L^{2})$%
-terms proportional to the field equations. Iterating in powers of $L$, we
find that the BR replacement has the form 
\begin{eqnarray*}
\varphi _{\mathrm{B}} &=&\varphi _{\mathrm{B}}(\varphi ,\lambda ,\mu
,L)=\varphi _{\mathrm{B}}(\varphi ,\lambda ,\mu )+\mathcal{O}(L),\qquad
\lambda _{\mathrm{B}}=\lambda _{\mathrm{B}}(\lambda ,\mu ),\qquad
\!\!\!\!\!\!\int J_{\mathrm{B}}\varphi _{\mathrm{B}}\leftrightarrow \int
J\varphi , \\
L_{I\mathrm{B}} &=&L_{J}(Z^{-1})_{I}^{J},\qquad \tau _{\mathrm{B}}=\tilde{%
\tau}(\tau ,\lambda ,\mu )Z^{-1}.
\end{eqnarray*}
The renormalized action is 
\begin{equation}
S_{L}(\varphi ,\lambda ,\mu ,L)=S(\varphi ,\lambda ,\mu ,L)-\int \left(
L_{I}+\hat{\tau}_{vI}\mathcal{N}^{v}(L,\lambda ,\mu )\right) \mathcal{O}_{%
\mathrm{R}}^{I}(\varphi ,\lambda ,\mu ),  \label{esln4}
\end{equation}
where $S(\varphi ,\lambda ,\mu ,L)=S_{\mathrm{B}}(\varphi _{\mathrm{B}%
},\lambda _{\mathrm{B}})$, and the parameter redefinitions $\tilde{\tau}%
_{vI} $ and $\hat{\tau}_{vI}$ need not coincide with those of (\ref{esln})
and (\ref{blareduline}).

Now we consider the BR change of variables. It is easy to see that $L_{I%
\mathrm{B}}$ cannot remain linear in $L$ and $J$. Indeed, when we make the
change of variables $\varphi _{\mathrm{B}}=\varphi _{\mathrm{B}}(\varphi
,\lambda ,\mu ,L)$ in the functional integral the term $\int J_{\mathrm{B}%
}\varphi _{\mathrm{B}}$ generates objects that can be absorbed only if we
introduce terms similar to the $\Delta f_{J}$s of formula (\ref{bla}), which
do not depend on $L$ and $J$ in any simple way.

Thus, if we want to keep linearity in $L$ and $J$ we must use the redundant
approach, or the trick mentioned above, where, besides converting $c$, $Z$, $%
\tilde{c}$ and $\tilde{Z}$ into derivative-operators, the terms proportional
to the field equations are viewed as terms proportional to descendants plus
mass terms and perturbative corrections.

\section{Changes of field variables in the redundant approach}

\setcounter{equation}{0}

In this section and the next two we study the most general perturbative
changes of field variables. Because of its simplicity, we prefer to
concentrate on the redundant approach and drop the essential one.

We explain how a change of variables in the action $S$ is related to a
change of variables in the $Z$- and $W$-functionals, namely how it reflects
from the integrand to the result of the functional integration. We do not
study the change of variables inside the $\Gamma $-functional, because this
investigation requires further work, which we leave to a separate paper.

Predictivity is unaffected by a change of variables. More explicitly, if the
number of independent physical couplings that are necessary (together with
field redefinitions) to reabsorb divergences is finite in some variable
frame, it is finite in every other variable frame. However, the change of
field variables itself requires its own renormalization. We show that it is
related to the renormalization of composite fields and work out this
relation explicitly.

Since the composite fields $\mathcal{O}_{c}^{I}$ form a basis for the local
functions of $\varphi $ and its derivatives, classically the most general
perturbative change of variables can be written in the form 
\begin{equation}
\varphi ^{\prime }(\varphi )=\varphi +\sum_{I}b_{I}\mathcal{O}%
_{c}^{I}(\varphi ),  \label{choffv}
\end{equation}
where $b_{I}=\mathcal{O}(\delta ^{n_{I}-1})$. We treat it perturbatively in
the constants $b_{I}$, so the functional integration measure is invariant.
We define classical composite fields 
\begin{equation}
\mathcal{O}_{c}^{I\hspace{0.01in}\prime }(\varphi ^{\prime })=\mathcal{O}%
_{c}^{I}(\varphi (\varphi ^{\prime }))  \label{newd}
\end{equation}
for the new variables, so the inverse of (\ref{choffv}) can be simply
written as 
\begin{equation}
\varphi (\varphi ^{\prime })=\varphi ^{\prime }-\sum_{I}b_{I}\mathcal{O}%
_{c}^{I\hspace{0.01in}\prime }(\varphi ^{\prime }).  \label{choffvinverse}
\end{equation}

\paragraph{Essential variable frame\newline
}

A parameter $\zeta $ is called \textit{inessential} if the derivative of the
action with respect to $\zeta $ is proportional to the field equations \cite
{weinberg}. A\ convenient choice of variables is the one where the action $%
S(\varphi ,\lambda ,\mu )$ does not contain inessential parameters.
Perturbatively, we can require that the action does not contain terms
proportional to $\delta S_{\text{free}}/\delta \varphi $, such as $(-\Box
+m_{s}^{2})\varphi $, $(\partial \!\!\!\slash+m_{f})\psi $, etc., and their
derivatives, apart from the quadratic terms we are perturbing around. We
call this reference frame the \textit{essential variable frame}. It is
useful in some applications.

The essential variable frame is preserved by renormalization. It is easy to
prove this statement directly, but we can also use the derivation of the BR
replacement (\ref{BRrepla}) in the essential approach. As explained before,
an equivalent way to describe the renormalization of an interacting theory
with classical action $S_{c}$, at $L=0$, is to replace $S_{c}$ with the
free-field action $S_{\text{free}}$, consider the extended action $S_{Lc}$
of (\ref{SLB2}) and replace the sources $L$ with constants, such that $%
S_{Lc} $ gives back $S_{c}$. Then the renormalization of the theory is
described by formula (\ref{BRrepla}). All counterterms proportional to $%
\delta S_{\text{free}}/\delta \varphi $ are subtracted by the field
redefinition, so the structure of the action in the essential variable frame
is preserved.

\paragraph{Bare change of field variables\newline
}

We first work at the bare level, where the change of variables is simpler,
and later discuss the change of variables at the renormalized level. We
start with the redundant nonlinear approach, where the bare action (\ref{SLB}%
) is linear in the sources $L_{\mathrm{B}}$. The bare change of variables
coincides in form with the classical one (\ref{choffv}), so we write 
\begin{equation}
\varphi _{\mathrm{B}}^{\prime }(\varphi _{\mathrm{B}})=\varphi _{\mathrm{B}%
}+b_{I\mathrm{B}}\mathcal{O}_{\mathrm{B}}^{I}(\varphi _{\mathrm{B}}).
\label{Bchoffv}
\end{equation}
Making the transformation 
\begin{equation}
L_{I\mathrm{B}}^{\prime }=L_{I\mathrm{B}}-b_{I\mathrm{B}}J_{\mathrm{B}%
},\qquad J_{\mathrm{B}}^{\prime }=J_{\mathrm{B}},  \label{blabla}
\end{equation}
and the change of variables (\ref{Bchoffv}) inside (\ref{ZLB}) we get 
\begin{equation}
Z_{\mathrm{B}}(J_{\mathrm{B}},L_{\mathrm{B}})=\int [\mathrm{d}\varphi _{%
\mathrm{B}}^{\prime }]\hspace{0.02in}\exp \left( -S_{L\mathrm{B}}^{\prime
}(\varphi _{\mathrm{B}}^{\prime },L_{\mathrm{B}}^{\prime })+\int \varphi _{%
\mathrm{B}}^{\prime }J_{\mathrm{B}}^{\prime }\right) ,  \label{zjbj}
\end{equation}
where 
\begin{equation}
S_{L\mathrm{B}}^{\prime }(\varphi _{\mathrm{B}}^{\prime },L_{\mathrm{B}%
}^{\prime })=S_{\mathrm{B}}^{\prime }(\varphi _{\mathrm{B}}^{\prime
})-\sum_{I}\int L_{\mathrm{B}}^{\prime \hspace{0.01in}I}\mathcal{O}_{\mathrm{%
B}}^{\prime \hspace{0.01in}I}(\varphi _{\mathrm{B}}^{\prime }),
\label{orario}
\end{equation}
and 
\[
S_{\mathrm{B}}^{\prime }(\varphi _{\mathrm{B}}^{\prime })=S_{\mathrm{B}%
}(\varphi _{\mathrm{B}}(\varphi _{\mathrm{B}}^{\prime })). 
\]
Observe that (\ref{zjbj}) coincides with the transformed bare functional $Z_{%
\mathrm{B}}^{\prime }(J_{\mathrm{B}}^{\prime },L_{\mathrm{B}}^{\prime })$.
Thus the functionals $Z_{\mathrm{B}}$ and $W_{\mathrm{B}}$ correctly behave
as scalars: 
\begin{equation}
Z_{\mathrm{B}}^{\prime }(J_{\mathrm{B}}^{\prime },L_{\mathrm{B}}^{\prime
})=Z_{\mathrm{B}}(J_{\mathrm{B}},L_{\mathrm{B}}),\qquad W_{\mathrm{B}%
}^{\prime }(J_{\mathrm{B}}^{\prime },L_{\mathrm{B}}^{\prime })=W_{\mathrm{B}%
}(J_{\mathrm{B}},L_{\mathrm{B}}).  \label{zpjbj2}
\end{equation}

As before, we can equivalently describe the map as the replacement 
\begin{equation}
\varphi _{\mathrm{B}}^{\prime }=\varphi _{\mathrm{B}}^{\prime }(\varphi _{%
\mathrm{B}}),\qquad L_{I\mathrm{B}}^{\prime }=L_{I\mathrm{B}},\qquad \int J_{%
\mathrm{B}}^{\prime }\varphi _{\mathrm{B}}^{\prime }\leftrightarrow \int J_{%
\mathrm{B}}\varphi _{\mathrm{B}}.  \label{Bchvrepla}
\end{equation}

\section{Renormalized changes of variables in the redundant approach}

\setcounter{equation}{0}

\label{rchov}

In this section we study the renormalized change of field variables in the
redundant nonlinear approach and show that it provides a simple method to
derive the renormalization of the theory in the new variables without having
to calculate it anew.

While a replacement, by definition, simply replaces the term $\int J\varphi $
with $\int J^{\prime }\varphi ^{\prime }$, a change of variables does
transform it as any other term, therefore switches the functional integral
from the conventional form to some unconventional one. We begin proving that
all perturbative $J$-dependencies besides the term $\int J\varphi $ can be
reabsorbed into a field redefinition, so it is always possible to rephrase
the functional integral in the conventional form. The proof of the theorem
also contains the procedure to achieve this result.

\paragraph{Switching from the non-conventional form to the conventional form%
\newline
}

\textbf{Theorem}. \textit{Consider a functional integral } 
\[
\mathcal{I}=\int [\mathrm{d}\varphi ]\hspace{0.02in}\exp \left( -S(\varphi
)+\int J\left( \varphi -bU\right) \right) , 
\]
\textit{where }$U(\varphi ,bJ)$\textit{\ is a local function of }$\varphi $%
\textit{\ and }$J$\textit{, and }$b$\textit{\ is a constant. Then there
exists a perturbatively local change of variables } 
\[
\varphi =\varphi (\varphi ^{\prime },b,bJ)=\varphi ^{\prime }+\mathcal{O}%
(b), 
\]
\textit{expressed as a series expansion in }$b$\textit{, such that } 
\[
\mathcal{I}=\int [\mathrm{d}\varphi ^{\prime }]\hspace{0.02in}\exp \left(
-S^{\prime }(\varphi ^{\prime },b)+\int J\varphi ^{\prime }\right) , 
\]
\textit{where} $S^{\prime }(\varphi ^{\prime },b)=S(\varphi (\varphi
^{\prime },b,0))$.

\textit{Proof}. Make the change of variables 
\begin{equation}
\varphi _{1}=\varphi -bU(\varphi ,bJ)  \label{phisech1}
\end{equation}
in the functional integral. The functional measure is invariant, since we
are treating (\ref{phisech1}) perturbatively in $b$. Call $\varphi
=f_{1}(\varphi _{1},b)$ the inverse of (\ref{phisech1}) at $J=0$. We can
write 
\[
S(\varphi )=S(f_{1}(\varphi _{1},b))+b^{2}\int JU_{1}, 
\]
for a suitable local function $U_{1}(\varphi _{1},bJ,b)$. Then we have 
\[
\mathcal{I}=\int [\mathrm{d}\varphi _{1}]\hspace{0.02in}\exp \left(
-S_{1}(\varphi _{1},b)+\int J\left( \varphi _{1}-b^{2}U_{1}\right) \right)
,\qquad S_{1}(\varphi _{1},b)=S(f_{1}(\varphi _{1},b)). 
\]

At this point, we are in the same situation we started with, but $U$ is
replaced by $bU_{1}$, which is one order of $b$ higher. Repeating the step
made above, we make the change of variables $\varphi _{2}=\varphi
_{1}-b^{2}U_{1}$ and get 
\[
\mathcal{I}=\int [\mathrm{d}\varphi _{2}]\hspace{0.02in}\exp \left(
-S_{2}(\varphi _{2},b)+\int J\left( \varphi _{2}-b^{3}U_{2}\right) \right)
,\qquad S_{2}(\varphi _{2},b)=S_{1}(f_{2}(\varphi _{2},b),b). 
\]
where $\varphi _{1}=f_{2}(\varphi _{2},b)$ is the inverse of $\varphi
_{2}=\varphi _{1}-b^{2}U_{1}$ at $J=0$ and $U_{2}(\varphi _{2},bJ,b)$ is a
local function. Proceeding indefinitely like this, we prove the theorem.

\paragraph{Renormalized change of variables\newline
}

In the remainder of this section we describe the renormalized change of
field variables working directly on renormalized quantities. The relation
between bare and renormalized changes of variables is worked out in section
11.

Start from the generating functional (\ref{ZLR}), which we write in the form 
\begin{equation}
Z(J,L)=\exp \left( \frac{1}{\hbar }W(J,L)\right) =\int [\mathrm{d}\varphi ]%
\hspace{0.02in}\exp \left( -\frac{1}{\hbar }S_{L}(\varphi ,L)+\int J\varphi
\right) ,  \label{changematch1}
\end{equation}
with the renormalized extended action (\ref{SLRredu}), where $%
f_{I}(L)=L_{I}+\Delta _{I}(L)$, $\Delta _{I}=\mathcal{O}(\hbar )=\mathcal{O}%
(L)$ and $S(\varphi )=S_{c}(\varphi )+\mathcal{O}(\hbar )$ is the
renormalized action, equal to the classical action $S_{c}$ plus its
counterterms. We have introduced $\hbar $ explicitly, because it is useful
for our argument. For the time being we omit the dependencies on $\lambda $
and $\mu $, since they are not crucial for the arguments that follow.

Now, shift the sources $L$ defining 
\begin{equation}
L_{I}^{\prime }=L_{I}-\hbar b_{I}J,  \label{mapl}
\end{equation}
and perform a change of variables 
\begin{equation}
\tilde{\varphi}(\varphi )=\varphi +b_{I}\mathcal{O}^{I}(\varphi ).
\label{choffv2}
\end{equation}
in the functional integral (\ref{changematch1}). We get 
\[
Z(J,L)=\int [\mathrm{d}\tilde{\varphi}]\hspace{0.02in}\exp \left( -\frac{1}{%
\hbar }\tilde{S}_{L}(\tilde{\varphi},\hbar bJ,L^{\prime })+\int J\tilde{%
\varphi}\right) , 
\]
where 
\[
\tilde{S}_{L}(\tilde{\varphi},\hbar bJ,L^{\prime })=S(\varphi (\tilde{\varphi%
}))-\int (L_{I}^{\prime }+\Delta _{I}(L^{\prime }+\hbar bJ))\mathcal{\tilde{O%
}}^{I}(\tilde{\varphi}), 
\]
and $\mathcal{\tilde{O}}^{I}(\tilde{\varphi})=\mathcal{O}^{I}(\varphi )$ is
the basis in the tilded variables. This result is not written in the
conventional form, yet, since $\tilde{S}_{L}$ depends on $J$. However, we
can use the theorem proved before to find the conventional form for the new
variables.

Specifically, we write 
\begin{equation}
\tilde{S}_{L}(\tilde{\varphi},\hbar bJ,L^{\prime })=\tilde{S}_{L}(\tilde{%
\varphi},0,L^{\prime })+\hbar ^{2}b\int JU(L^{\prime },\hbar bJ,\tilde{%
\varphi}),  \label{samesitu}
\end{equation}
for a suitable local function $U$, where $b$ collectively denotes the
parameters $b_{I}$. The second term of (\ref{samesitu}) is $\mathcal{O}%
(\hbar ^{2})$, because $\Delta ^{I}(L)=\mathcal{O}(\hbar )$. The generating
functional becomes 
\[
Z(J,L)=\int [\mathrm{d}\tilde{\varphi}]\hspace{0.02in}\exp \left( -\frac{1}{%
\hbar }\tilde{S}_{L}(\tilde{\varphi},0,L^{\prime })+\int J\left( \tilde{%
\varphi}-\hbar bU\right) \right) . 
\]
The theorem proved before ensures that there exists a perturbatively local
change of variables $\tilde{\varphi}=\tilde{\varphi}(\varphi ^{\prime
},J,L^{\prime })=\varphi ^{\prime }+\mathcal{O}(\hbar b)$ that converts the
functional integral to the conventional form, such that 
\[
Z(J,L)=\int [\mathrm{d}\varphi ^{\prime }]\hspace{0.02in}\exp \left( -\frac{1%
}{\hbar }\tilde{S}_{L}(\tilde{\varphi}(\varphi ^{\prime },0,L^{\prime
}),0,L^{\prime })+\int J\varphi ^{\prime }\right) . 
\]
Now it remains to expand $\tilde{\varphi}(\varphi ^{\prime },0,L^{\prime })$
in powers of $L^{\prime }$. Call 
\[
\mathcal{O}^{I\hspace{0.01in}\prime }(\varphi ^{\prime })=\mathcal{\tilde{O}}%
^{I}(\tilde{\varphi}(\varphi ^{\prime },0,0))=\mathcal{O}^{I}(\varphi (%
\tilde{\varphi}(\varphi ^{\prime },0,0))),\qquad S^{\prime }(\varphi
^{\prime })=S(\varphi (\tilde{\varphi}(\varphi ^{\prime },0,0))), 
\]
the new basis of composite fields and the new action, respectively. We can
find local $L^{\prime }$-dependent functions $s_{J}^{I}$ and $r^{I}$ such
that 
\[
\mathcal{O}^{I}(\varphi (\tilde{\varphi}(\varphi ^{\prime },0,L^{\prime
})))=s_{J}^{I}(L^{\prime })\mathcal{O}^{J\hspace{0.01in}\prime }(\varphi
^{\prime }),\qquad S(\varphi (\tilde{\varphi}(\varphi ^{\prime },0,L^{\prime
})))=S^{\prime }(\varphi ^{\prime })+\int r_{I}(L^{\prime })\mathcal{O}^{I%
\hspace{0.01in}\prime }(\varphi ^{\prime }). 
\]
Clearly, both $s_{J}^{I}-\delta _{J}^{I}$ and $r_{I}$ are $\mathcal{O}(\hbar
)$ and $\mathcal{O}(L^{\prime })$. Finally, defining 
\begin{equation}
S_{L}^{\prime }(\varphi ^{\prime },L^{\prime })=S^{\prime }(\varphi ^{\prime
})-\int f_{I}^{\prime }(L^{\prime })\mathcal{O}^{I\hspace{0.01in}\prime
}(\varphi ^{\prime }),  \label{nevech}
\end{equation}
where 
\begin{equation}
f_{I}^{\prime }(L^{\prime })=f_{J}(L^{\prime })s_{I}^{J}(L^{\prime
})-r_{I}(L^{\prime }),  \label{implu}
\end{equation}
the generating functional reads 
\begin{equation}
Z(J,L)=\int [\mathrm{d}\varphi ^{\prime }]\hspace{0.02in}\exp \left( -\frac{1%
}{\hbar }S_{L}^{\prime }(\varphi ^{\prime },L^{\prime })+\int J\varphi
^{\prime }\right) .  \label{ostri}
\end{equation}
The right-hand side of this formula is precisely the generating functional $%
Z^{\prime }(J,L^{\prime })$, as it is quantized and renormalized in the new
variables. We conclude that the change of field variables reads 
\begin{equation}
\varphi ^{\prime }=\varphi ^{\prime }(\varphi ,J,L^{\prime }),\qquad
J^{\prime }=J,\qquad L_{I}^{\prime }=L_{I}-\hbar b_{I}J,  \label{shift2r}
\end{equation}
where $\varphi ^{\prime }=\varphi ^{\prime }(\varphi ,J,L^{\prime })$ is the
inverse of $\varphi =\varphi (\tilde{\varphi}(\varphi ^{\prime },J,L^{\prime
}))$, and the $Z$- and $W$-functionals behave as scalars: 
\begin{equation}
Z^{\prime }(J^{\prime },L^{\prime })=Z(J,L),\qquad W^{\prime }(J^{\prime
},L^{\prime })=W(J,L).  \label{mapzw}
\end{equation}
Note that formula (\ref{implu}) encodes the relation between the
renormalizations of composite fields before and after the change of field
variables.

We also have 
\begin{equation}
S_{L}^{\prime }(\varphi ^{\prime },L^{\prime })=\tilde{S}_{L}(\tilde{\varphi}%
(\varphi ^{\prime },0,L^{\prime }),0,L^{\prime })=S_{L}(\varphi (\tilde{%
\varphi}(\varphi ^{\prime },0,L^{\prime })),L^{\prime }).  \label{alsohave}
\end{equation}
Ultimately, the change of field variables has three aspects: $i$)\ in the
functional integral we make the change of integration variables $\varphi
^{\prime }=\varphi ^{\prime }(\varphi ,J,L^{\prime })$; $ii$) inside the
extended action $S_{L}$ we have the change of variables $\varphi ^{\prime
}=\varphi ^{\prime }(\varphi ,0,L^{\prime })$; $iii$) inside the action $%
S(\varphi )$ we just have $\varphi ^{\prime }=\varphi ^{\prime }(\varphi
,0,0)$.

Summarizing, when we make the change of variables (\ref{choffv2}) we get
unwanted $J$-dependent terms from $\int J\varphi $. We cancel them by means
of the source redefinitions (\ref{mapl}). However, (\ref{mapl}) generate
other unwanted $J$-dependent terms. Those are canceled upgrading the $J$%
-independent change of variables (\ref{choffv2}) to a $J$-dependent one,
which is $\varphi ^{\prime }=\varphi ^{\prime }(\varphi ,J,L^{\prime })$.

All transformations $i$), $ii$) and $iii$) are equal to (\ref{choffv2}) plus
appropriate counterterms. A change of variables undergoes its own
renormalization, which is related to the renormalization of the composite
fields it is made of. The derivation just given also teaches us how to work
it out.

The most general local redefinitions of $L$ can be considered, instead of
those of (\ref{shift2r}). They amount to combinations of changes of
variables and redefinitions of the basis $\mathcal{O}^{I}$. Relations (\ref
{blaredu}) show that renormalization is a redefinition of this more general
type, in the redundant nonlinear approach. The change of field variables is
always encoded inside the $J$-dependence of the $L$-redefinitions.
Redefinitions of $J$, instead, are never necessary, since the elementary
field is also included in the basis $\mathcal{O}^{I}$.

As before, we can describe the effects of the change of variables with a
replacement. We can actually give two equivalent forms of the replacement.
Using (\ref{alsohave}) we can make 
\begin{equation}
\varphi ^{\prime }=\varphi ^{\prime }(\varphi ,0,L^{\prime }),\qquad
L^{\prime }=L,\qquad \int J^{\prime }\varphi ^{\prime }\leftrightarrow \int
J\varphi .  \label{replo}
\end{equation}
Alternatively, we can make 
\begin{equation}
\varphi ^{\prime }=\varphi ^{\prime }(\varphi ,0,0),\qquad L^{\prime
}=L^{\prime }(L),\qquad \int J^{\prime }\varphi ^{\prime }\leftrightarrow
\int J\varphi ,  \label{chvrepla}
\end{equation}
where $L^{\prime }(L)$ are the solutions of 
\[
f_{I}(L)=f_{I}^{\prime }(L^{\prime })=f_{J}(L^{\prime })s_{I}^{J}(L^{\prime
})-r_{I}(L^{\prime }), 
\]
which can be worked out perturbatively in $\hbar $.

Combining (\ref{BRreplaredu}), (\ref{Bchvrepla}) and (\ref{chvrepla}) we can
read the primed BR replacement 
\begin{equation}
\varphi _{\mathrm{B}}^{\prime }=\varphi _{\mathrm{B}}^{\prime }(\varphi _{%
\mathrm{B}}(\varphi (\varphi ^{\prime }))),\qquad \lambda _{\mathrm{B}%
}=\lambda _{\mathrm{B}}(\lambda ,\mu ),\qquad L_{I\mathrm{B}}^{\prime
}=f_{I}^{\prime }(L^{\prime }),\qquad \!\!\!\!\!\!\int J_{\mathrm{B}%
}^{\prime }\varphi _{\mathrm{B}}^{\prime }\leftrightarrow \int J^{\prime
}\varphi ^{\prime },  \label{BRreplareduprime}
\end{equation}
where the function $\varphi (\varphi ^{\prime })$ is the inverse of $\varphi
^{\prime }=\varphi ^{\prime }(\varphi ,0,0)$.

Note that the functions $\varphi (\varphi ^{\prime })$ and $L^{\prime }(L)$
are in general divergent, yet the generating functionals remain convergent,
because they are mapped into each other by the convergent relations (\ref
{mapl}) and $J^{\prime }=J$. The divergences contained in $\varphi (\varphi
^{\prime })$ and $L^{\prime }(L)$ are the extra counterterms necessary to
turn the renormalization of the theory expressed in the old variables into
the renormalization of the theory expressed in the new variables.

However, the replacement (\ref{chvrepla})\ is just a merely descriptive
existence relation between the renormalizations in the old and the new
variable frames. It is not equipped with a method to calculate the functions 
$\varphi ^{\prime }=\varphi ^{\prime }(\varphi ,0,0)$ and $L^{\prime
}=L^{\prime }(L)$. The only ways we have to derive those functions are
either using the change of variables or calculating Feynman diagrams anew in
the new variables.

In other words, if we do not want to recalculate Feynman diagrams from
scratch, we just apply the classical variable change and then recover the
conventional form of the functional integral following the steps described
in this section. Thus the change of variables provides an independent way to
derive the renormalization of the theory in the new variables from its
renormalization in the old variables.

The explanation of this crucial difference between replacements and true
changes of field variables is that only changes of variables take full
advantage of composite fields. The role of composite fields in the
replacement (\ref{chvrepla}) is minor, to the extent that they can be freely
switched on and off in (\ref{chvrepla}) with practically no gain nor loss.

In the new variables the renormalization program works as in the old
variables. The renormalization of parameters remains the same, but the field
renormalizations can change considerably. For example, if the fields
renormalize multiplicatively before the change of variables, or do not
renormalize at all, they may renormalize non-polynomially after the change
of variables. Nevertheless, a theory that is predictive in some variables
(which means that it can be renormalized with redefinitions of a finite
number of physical parameters and local field redefinitions), remains
predictive in any other variable frame. This ensures that the physics
remains the same. In section 12 we give explicit examples.

The source redefinitions (\ref{shift2r}) are linear and encode the most
general changes of field variables. However, we point out that the
functional $W^{\prime }(J^{\prime },L^{\prime })\equiv W(J,L)$ obtained
applying any perturbatively local source redefinitions 
\begin{equation}
J^{\prime }=J,\qquad L^{\prime }=L^{\prime }(J,L)=L^{\prime }+\mathcal{O}(b),
\label{chW}
\end{equation}
with $L^{\prime }(0,0)=0$, is the $W$-functional that we would calculate in
some transformed field-variable frame. The transformed fields can be worked
out applying the procedure explained in this section to recover the
conventional form of the functional integral, which is spoiled by nontrivial 
$J$-dependences contained in $L^{\prime }(J,L)$.

\section{Changes of field variables in the linear approach}

\setcounter{equation}{0}

Now we examine the change of variables in the linear approach of section 7.
Equipped with the experience already gained, this task is now relatively
easy. We can go back to work with $\hbar =1$.

Recall that the classical action is (\ref{bibop}) and the renormalized one
is (\ref{esln}). Make the substitutions (\ref{mapl}) in (\ref{esln}), where $%
b_{I}=\mathcal{O}(\delta ^{n_{I}-1})$. Such substitutions certainly leave
the generating functionals $Z$ and $W$ convergent, but do not preserve the
conventional form of the functional integral. We just have to convert the $Z$%
-integrand back to the conventional form. Then we can read the change of
variables associated with (\ref{mapl}) and the renormalization constants $%
Z^{\prime }=Z+\mathcal{O}(b)$ and $\hat{\tau}^{\prime }=\hat{\tau}+\mathcal{O%
}(b)$ that remove the divergences in the new variables.

To do this, we make the change of variables 
\begin{equation}
\tilde{\varphi}(\varphi )=\varphi +\left( b_{I}+\Delta b_{I}\right)
(Z^{-1})_{J}^{I}\mathcal{O}^{J}(\varphi ),  \label{blabla2}
\end{equation}
in the functional integral, instead of (\ref{choffv2}), where $\Delta b_{I}=%
\mathcal{O}(b^{2})$ are constants to be determined. It may be convenient to
express (\ref{blabla2}) as 
\[
\varphi (\tilde{\varphi})=\tilde{\varphi}-\left( b_{I}+\Delta b_{I}\right)
(Z^{-1})_{J}^{I}\mathcal{\tilde{O}}^{J}(\tilde{\varphi}), 
\]
where $\mathcal{\tilde{O}}^{I}(\tilde{\varphi})=\mathcal{O}^{I}(\varphi )$.
The exponent of the $Z$-integrand can be written as 
\begin{equation}
-S(\varphi (\tilde{\varphi}))+\int \left( L_{I}^{\prime }-\Delta b_{I}J+\hat{%
\tau}_{vI}\mathcal{N}^{v}(L^{\prime })\right) (Z^{-1})_{J}^{I}\mathcal{%
\tilde{O}}^{J}(\tilde{\varphi})+\int J(\tilde{\varphi}+b_{I}U^{I}),
\label{www}
\end{equation}
where $U^{I}(\tilde{\varphi},bJ,L^{\prime },b+\Delta b)$ are $\mathcal{O}(%
\tilde{L}^{\prime })$-local functions of $\tilde{L}^{\prime }=\{L^{\prime
},J\}$. Now we show that $\Delta b$ can be determined perturbatively in $b$
so that (\ref{www}) is converted to the conventional form.

Make the further change of variables $\tilde{\varphi}\rightarrow \tilde{%
\varphi}-b_{I}U^{I}$. The new action can be expanded as 
\begin{eqnarray}
S(\varphi (\tilde{\varphi})) &\rightarrow &S(\varphi (\tilde{\varphi}))+\int
\left( L_{J}^{\prime }\bar{\Delta}z_{I}^{J}-J\bar{\Delta}b_{I}+\bar{\Delta}%
\hat{\tau}_{vI}\mathcal{N}^{v}(L^{\prime })\right) (Z^{-1})_{K}^{J}\mathcal{%
\tilde{O}}^{K}(\tilde{\varphi})  \nonumber \\
&&-b^{2}\int JU^{\prime }(\tilde{\varphi},bJ,L^{\prime },b+\Delta b,b),
\label{erto2}
\end{eqnarray}
where $\bar{\Delta}z\sim b\hat{\Delta}z$, $\bar{\Delta}b\sim b^{2}\hat{\Delta%
}b$ and $\bar{\Delta}\hat{\tau}\sim b\hat{\Delta}\hat{\tau}$, where $\hat{%
\Delta}z$, $\hat{\Delta}b$ and $\hat{\Delta}\hat{\tau}$ are functions of $b$
and $b+\Delta b$, while $U^{I}$ are $\mathcal{O}(\tilde{L}^{\prime })$-local
functions. Now, determine $\Delta b_{I}$ to $\mathcal{O}(b^{2})$ so that $%
\Delta b_{I}=\bar{\Delta}b_{I}+\mathcal{O}(b^{3})$ and define $Z^{\prime
}=Z(1+\bar{\Delta}z)+\mathcal{O}(b^{2})$. An expansion similar to (\ref
{erto2}) can be written for the transformed $\mathcal{\tilde{O}}^{I}(\tilde{%
\varphi})$s, and generates additional terms like the last two of (\ref{erto2}%
). Terms like the last-but-one of (\ref{erto2}) determine the $\mathcal{O}%
(b) $-corrections to $\hat{\tau}$ that define $\hat{\tau}^{\prime }$. Terms
like the last one of (\ref{erto2}) have the same structure as the term $\int
JbU$ of (\ref{www}), but are one order higher in $b$. Repeating the
procedure indefinitely, the complete change of variables in the functional
integral gets of the form 
\[
\varphi ^{\prime }=\varphi ^{\prime }(\varphi ,J,L)=\varphi +\left(
b_{I}+\Delta b_{I}\right) (Z^{\hspace{0.01in}-1})_{J}^{I}\mathcal{O}%
^{J}(\varphi )+b\hspace{0.01in}_{I}\hat{U}^{I}(\varphi ,bJ,L,b),\qquad \hat{U%
}^{I}=\mathcal{O}(\tilde{L}), 
\]
and the exponent of the $Z$-integrand turns into its correct primed version,
which is 
\begin{equation}
-S^{\prime }(\varphi ^{\prime })+\int \left( L_{J}^{\prime }+\hat{\tau}%
_{vJ}^{\prime }\mathcal{N}^{v}(L^{\prime })\right) (Z^{\prime \hspace{0.01in}%
-1})_{I}^{J}\mathcal{O}^{I\hspace{0.01in}\prime }(\varphi ^{\prime })+\int
J^{\prime }\varphi ^{\prime }=-S_{L}^{\prime }(\varphi ^{\prime },L^{\prime
})+\int J^{\prime }\varphi ^{\prime },  \label{terry}
\end{equation}
where $S^{\prime }(\varphi ^{\prime })=S(\varphi (\varphi ^{\prime },0,0))$
and $\mathcal{O}^{I\hspace{0.01in}\prime }(\varphi ^{\prime })=\mathcal{O}%
^{I}(\varphi (\varphi ^{\prime },0,0))$, while $J^{\prime }=J$, $Z^{\prime
}=Z+\mathcal{O}(b)$ and $\hat{\tau}^{\prime }=\hat{\tau}+\mathcal{O}(b)$.
Thus, the renormalized change of variables in the linear redundant approach
still has the form (\ref{shift2r}). The renormalized composite fields
transform as 
\[
\mathcal{O}_{\mathrm{R}}^{\prime \hspace{0.01in}I}(\varphi ^{\prime
})=(Z^{\prime \hspace{0.01in}-1})_{I}^{J}\mathcal{O}^{\prime \hspace{0.01in}%
I}(\varphi ^{\prime })=(Z^{\prime \hspace{0.01in}-1}Z)_{I}^{J}\mathcal{O}_{%
\mathrm{R}}^{I}(\varphi ). 
\]

We can also view the change of variables as the replacement 
\[
\varphi ^{\prime }=\varphi ^{\prime }(\varphi ,0,0),\qquad L_{I}^{\prime
}=L_{J}(Z^{-1}Z^{\prime })_{I}^{J},\qquad \hat{\tau}_{vI}^{\prime
}=(C^{-1})_{v}^{w}\hat{\tau}_{wJ}(Z^{-1}Z^{\prime })_{I}^{J},\qquad \int
J^{\prime }\varphi ^{\prime }\leftrightarrow \int J\varphi , 
\]
where $C$ is the matrix such that $\mathcal{N}^{v}(L^{\prime })=C_{w}^{v}%
\mathcal{N}^{w}(L)$. Again, the replacement is a mere description of the
result, because it does not provide an independent way to calculate the
quantities appearing in the transformation. Instead, the operations
described with the change of integration variables do allow us to derive the
renormalization of the transformed theory from the renormalization of the
original one, without having to calculate diagrams from scratch in the new
variables.

We learn that a finite change of variables can always be expressed with a
linear source redefinition of the form (\ref{mapl}), both in the linear and
nonlinear approaches. On the other hand, the BR map, which includes a
divergent change of field variables, can be expressed as a linear source
redefinition only in the linear approach, whence the name we have given to
this approach.

Let us analyze the result we have obtained in more detail. If the theory is
renormalized using the minimal subtraction scheme in the variables $\varphi $%
, in general it will not be renormalized using the minimal subtraction
scheme after the change of variables. The reason is that if, for example, $%
Z=1+$poles in $\varepsilon =4-D$, where $D$ is the continued dimension in
the dimensional regularization, $Z^{\prime }$ needs not be equal to $1+$%
poles. Nevertheless, we can extract the finite part writing $Z^{\prime }=%
\bar{Z}^{\prime }\bar{z}$, where $\bar{z}=1+\mathcal{O}(b)$ is finite and $%
\bar{Z}^{\prime }=1+$poles. Similarly, although $\hat{\tau}=\tau +$poles, $%
\hat{\tau}^{\prime }$ is equal to some finite function $\tau ^{\prime }(\tau
,b,\lambda ,\mu )=\tau +\mathcal{O}(b)$ plus poles. We can view $\tau
^{\prime }$ as a redefinition of $\tau $. There also exist finite
redefinitions $\tilde{b}_{I}(b,\lambda ,\tau ,\mu )=b_{I}+\mathcal{O}(b^{2})$
such that 
\begin{equation}
\varphi ^{\prime }(\varphi ,0,0)=\varphi +\tilde{b}_{I}(b,\lambda ,\tau ,\mu
)\mathcal{O}_{c}^{I}(\varphi )+\text{poles.}  \label{relacus}
\end{equation}

We can preserve the minimal subtraction scheme if we include a finite change
of basis $L_{I}^{\prime }\rightarrow L_{J}^{\prime }\bar{z}_{I}^{J}$. In
other words, instead of (\ref{mapl}) we define the source redefinitions as 
\[
L_{I}^{\prime }=(L_{J}-b_{J}J)(\bar{z}^{-1})_{I}^{J},\qquad J^{\prime }=J. 
\]
Then the $S_{L}^{\prime }$-terms linear in $L^{\prime }$ are $\int
L_{J}^{\prime }(\bar{Z}^{\prime \hspace{0.01in}-1})_{I}^{J}\mathcal{O}%
^{\prime \hspace{0.01in}I}(\varphi ^{\prime })$, and the transformed
renormalized composite fields are 
\begin{equation}
\mathcal{O}_{\mathrm{R}}^{\prime \hspace{0.01in}I}(\varphi ^{\prime })=(\bar{%
Z}^{\prime \hspace{0.01in}-1})_{I}^{J}\mathcal{O}^{\prime \hspace{0.01in}%
I}(\varphi ^{\prime })=(\bar{Z}^{\prime \hspace{0.01in}-1}Z)_{I}^{J}\mathcal{%
O}_{\mathrm{R}}^{I}(\varphi ).  \label{newd2}
\end{equation}

More generally, we can always include a further finite change of basis $%
L_{I}^{\prime }\rightarrow L_{J}^{\prime }\tilde{z}_{I}^{J}$, $\mathcal{O}_{%
\mathrm{R}}^{\prime \hspace{0.01in}I}\rightarrow (\tilde{z}^{-1})_{J}^{I}%
\mathcal{O}_{\mathrm{R}}^{\prime \hspace{0.01in}J}$ and describe the change
of variables as the more general map 
\begin{eqnarray}
\varphi ^{\prime } &=&\varphi ^{\prime }(\varphi ,J,L),\qquad b^{\prime
}=b^{\prime }(b,\lambda ,\tau ,\mu ),\qquad \tau ^{\prime }=\tau ^{\prime
}(\tau ,b,\lambda ,\mu ),  \nonumber \\
L_{I}^{\prime } &=&(L_{J}-b_{J}J)(z^{-1})_{I}^{J},\qquad J^{\prime }=J,
\label{blareduline2}
\end{eqnarray}
where $z=\tilde{z}\bar{z}$ and the primed parameters $b_{I}^{\prime
}(b,\lambda ,\tau ,\mu )=b_{I}+\mathcal{O}(b^{2})$ are finite functions
obtained inverting the relation (\ref{relacus}) so that it reads 
\begin{equation}
\varphi (\varphi ^{\prime },0,0)=\varphi ^{\prime }-b_{I}^{\prime }\mathcal{O%
}_{c}^{\prime \hspace{0.01in}I}(\varphi ^{\prime })+\text{poles.}
\label{treeck}
\end{equation}
The finite functions $b^{\prime }$, $\tau ^{\prime }$ and $z$ can be chosen
to combine the change of field variables with any change of subtraction
scheme in the composite-field sector. In particular they can be chosen to
preserve the minimal subtraction scheme.

We have already stressed that the virtue of the linear approach is that it
linearizes the map relating bare and renormalized sources. Thanks to this
fact, the procedure to make a bare change of field variables is practically
identical to the one just described at the renormalized level. We just
present it quickly and report the result. We start from (\ref{SLBlinear})
and make the substitutions 
\[
L_{I\mathrm{B}}^{\prime }=(L_{J\mathrm{B}}-b_{J\mathrm{B}}J_{\mathrm{B}})(z_{%
\mathrm{B}}^{-1})_{I}^{J},\qquad \tau _{\mathrm{B}}=\tau _{\mathrm{B}%
}^{\prime }+\Delta \tau _{\mathrm{B}}, 
\]
and a change of variables 
\[
\tilde{\varphi}_{\mathrm{B}}(\varphi _{\mathrm{B}})=\varphi _{\mathrm{B}%
}+(b_{I\mathrm{B}}+\Delta b_{I\mathrm{B}})\mathcal{O}_{\mathrm{B}%
}^{I}(\varphi _{\mathrm{B}}), 
\]
where $(z_{\mathrm{B}})_{I}^{J}=\delta _{I}^{J}+\mathcal{O}(b)$, $\Delta b_{I%
\mathrm{B}}=\mathcal{O}(b^{2})$ and $\Delta \tau _{\mathrm{B}}=\mathcal{O}%
(b) $. We obtain an expression similar to (\ref{www}). Then we make the
further change of variables $\tilde{\varphi}_{\mathrm{B}}\rightarrow \tilde{%
\varphi}_{\mathrm{B}}-b_{I\mathrm{B}}U_{\mathrm{B}}^{I}$, expand the new
action and the composite fields as in (\ref{erto2}), and determine the first
contributions to $(z_{\mathrm{B}})_{I}^{J}-\delta _{I}^{J}$, $\Delta b_{I%
\mathrm{B}}$ and $\Delta \tau _{\mathrm{B}}$. Repeating this procedure
indefinitely we arrive at the primed version of (\ref{SLBlinear}). Finally,
the renormalized change of variables in the linear redundant approach has
the form 
\begin{eqnarray}
\varphi _{\mathrm{B}}^{\prime } &=&\varphi _{\mathrm{B}}^{\prime }(\varphi _{%
\mathrm{B}},J_{\mathrm{B}},L_{\mathrm{B}})=\tilde{\varphi}_{\mathrm{B}%
}(\varphi _{\mathrm{B}})+b_{\mathrm{B}}\hspace{0.01in}\hat{U}_{\mathrm{B}%
}(\varphi _{\mathrm{B}},b_{\mathrm{B}}J_{\mathrm{B}},L_{\mathrm{B}}),\qquad
J_{\mathrm{B}}^{\prime }=J_{\mathrm{B}},  \nonumber \\
L_{I\mathrm{B}}^{\prime } &=&(L_{J\mathrm{B}}-b_{J\mathrm{B}}J_{\mathrm{B}%
})(z_{\mathrm{B}}^{-1})_{I}^{J},\qquad \tau _{\mathrm{B}}^{\prime }=\tau _{%
\mathrm{B}}+\mathcal{O}(b),  \label{blareduline3}
\end{eqnarray}
and $\hat{U}_{\mathrm{B}}=\mathcal{O}_{\mathrm{B}}(\tilde{L}_{\mathrm{B}})$.
The basis of composite fields inherits some change of basis $\mathcal{O}_{%
\mathrm{B}}^{\prime \hspace{0.01in}I}(\varphi _{\mathrm{B}}^{\prime })=(%
\tilde{z}_{\mathrm{B}}^{-1})_{J}^{I}\mathcal{O}_{\mathrm{B}}^{J}(\varphi _{%
\mathrm{B}})$, with $\tilde{z}_{\mathrm{B}}=z_{\mathrm{B}}+\mathcal{O}(b)$.

\section{Relation between bare and renormalized changes of variables}

\setcounter{equation}{0}

Having expressed the BR map as a change of variables, now it is simple to
work out the relation between bare and renormalized changes of variables. We
want to close the scheme 
\begin{equation}
\begin{tabular}{ccc}
$\mathrm{B}$ & $\leftrightarrow $ & $\mathrm{R}$ \\ 
$\updownarrow $ &  & $\updownarrow $ \\ 
$\mathrm{B}^{\prime }$ & $\leftrightarrow $ & $\mathrm{R}^{\prime }$%
\end{tabular}
\label{schiena}
\end{equation}
which gives us another way to express the map $\mathrm{R}\leftrightarrow 
\mathrm{R}^{\prime }$ and clarifies some points.

We start from the nonlinear approach. Composing the renormalized change of
variables (\ref{blaredu}) with (\ref{blabla}) and the primed analogue of (%
\ref{blaredu}), we obtain $J^{\prime }=J$ and 
\begin{equation}
f_{I}^{\prime }(L^{\prime })=f_{I}(L)+(c_{I}^{\prime }-c_{I}-b_{I\mathrm{B}%
})J.  \label{afgi}
\end{equation}
It is not evident how these relations can be compatible with (\ref{shift2r}%
). Using (\ref{implu}) we can solve (\ref{afgi}) to find $L^{\prime }$ as
functions of $L$ and $J$. However, we certainly obtain a divergent
transformation rule for the renormalized sources, not a relation of the form
(\ref{shift2r}). Yet, the generating functionals are convergent and these
maps must be equivalent to (\ref{shift2r}).

The matter can be explained as follows. The dependencies on $J$ and the $L$%
s\ are related to each other, because composite fields are ultimately made
of elementary fields. For example, if the bare action is written in the form
(\ref{SLB}) we can write 
\begin{equation}
\frac{\delta W_{\mathrm{B}}}{\delta L_{I\mathrm{B}}}=\langle \mathcal{O}_{%
\mathrm{B}}^{I}(\varphi _{\mathrm{B}})\rangle =\frac{1}{Z_{\mathrm{B}}}%
\mathcal{O}_{\mathrm{B}}^{I}(\frac{\delta }{\delta J_{\mathrm{B}}})Z_{%
\mathrm{B}}.  \label{aias}
\end{equation}
Similarly, if $L_{2\mathrm{B}}$ and $L_{4\mathrm{B}}$ are the sources
coupled with $\varphi _{\mathrm{B}}^{2}/2$ and $\varphi _{\mathrm{B}}^{4}/4!$%
, respectively, then 
\begin{equation}
\frac{\delta W_{\mathrm{B}}^{{}}}{\delta L_{4\mathrm{B}}}=\frac{1}{4!}%
\langle \varphi _{\mathrm{B}}^{4}\rangle =\frac{1}{6}\left\langle \left( 
\frac{\varphi _{\mathrm{B}}^{2}}{2}\right) ^{2}\right\rangle =\frac{1}{6}%
\frac{\delta ^{2}W_{\mathrm{B}}^{{}}}{\delta L_{2\mathrm{B}}^{2}}+\frac{1}{6}%
\left( \frac{\delta W_{\mathrm{B}}^{{}}}{\delta L_{2\mathrm{B}}}\right) ^{2}.
\label{aias2}
\end{equation}
At the renormalized level, these identities may get corrections that
compensate for the divergences originating from $J$- and $L$-derivatives at
coinciding points. We call identities like (\ref{aias}), (\ref{aias2}) and
their renormalized counterparts \textit{secret identities}. Due to the
secret identities, there exist convergent as well as divergent redefinitions
of $J$ and $L$ that leave the generating functionals invariant. In the sect
section we give an explicit example.

If some divergent $J$-$L$ redefinitions leave the renormalized functionals
convergent, we can find equivalent convergent redefinitions dropping the
divergent parts. In the minimal subtraction scheme, where bare and
renormalized quantities differ by pure poles, it is sufficient to drop the
pole corrections. Secret identities ensure that the poles of such $J$-$L$
redefinitions have no effect on the functionals. It is easy to see that the
function $U$ of (\ref{samesitu}) is divergent, therefore $s_{I}^{J}=\delta
_{I}^{J}+$poles, $r_{I}=$poles. Moreover, $c_{I}=$poles and $f_{I}(L)=L_{I}+$%
poles. Thus, dropping all divergent corrections from the relations (\ref
{afgi}) we get precisely (\ref{shift2r}). In another subtraction scheme the
finite equivalent versions of (\ref{afgi}) are, in general, a combination of
(\ref{shift2r}) with a scheme change.

Observe that the secret identities like (\ref{aias}) and (\ref{aias2}) are
nonlinear in the derivatives with respect to the sources. Thus, the problem
just described is absent in the linear approach. There the maps relating the
sources $L$ and $J$ are always linear, so their composition is still linear.
Composing (\ref{blareduline}), (\ref{blareduline3}) and the primed version
of (\ref{blareduline}), we get $J^{\prime }=J$, 
\begin{equation}
L_{I}^{\prime }=(L\tilde{Z}^{-1}z_{\mathrm{B}}^{-1}\tilde{Z}^{\prime
})_{I}+\left[ \tilde{c}_{I}^{\prime }-(\tilde{c}\tilde{Z}^{-1}+b_{\mathrm{B}%
})_{J}(z_{\mathrm{B}}^{-1}\tilde{Z}^{\prime })_{I}^{J}\right] J,
\label{tuitt}
\end{equation}
and some maps $\tau ^{\prime }=\tilde{\tau}^{\prime }(\tau ,\lambda ,\mu )$, 
$\varphi ^{\prime }=F(\varphi ,\lambda ,\mu ,J,L)$. Comparing (\ref{tuitt})
and (\ref{blareduline2}), we find 
\[
z_{\mathrm{B}}=\tilde{Z}^{\prime }z\tilde{Z}^{-1},\qquad b_{\mathrm{B}}=(b-%
\tilde{c}+\tilde{c}^{\prime }z)\tilde{Z}^{-1}. 
\]

\section{Examples}

In this section we collect a number of examples that illustrate various
properties derived in the paper.

\paragraph{Example 1\newline
}

Consider the classical theory 
\[
S_{c}(\varphi )=\frac{1}{2}\int \mathrm{d}^{D}x\hspace{0.01in}\left( 1+\frac{%
\lambda ^{2}}{2}\varphi ^{2}\right) (\partial _{\mu }\varphi )^{2}. 
\]
This is a derivative $\varphi ^{4}$-theory. It is equivalent to the massless
free scalar field up to the change of variables 
\begin{equation}
\varphi _{c}^{\prime }(\varphi )=\frac{\varphi }{2}\sqrt{1+\frac{\lambda ^{2}%
}{2}\varphi ^{2}}+\frac{1}{\lambda \sqrt{2}}\text{arcsinh}\left( \frac{%
\lambda \varphi }{\sqrt{2}}\right) .  \label{classchv}
\end{equation}
Indeed, 
\[
S_{c}(\varphi )=\frac{1}{2}\int \mathrm{d}^{D}x\hspace{0.01in}(\partial
_{\mu }\varphi _{c}^{\prime }(\varphi ))^{2}\equiv S_{c}^{\prime }(\varphi
_{c}^{\prime }). 
\]
We want to study the renormalization of $S_{c}(\varphi )$ at one loop and
verify that in the new variable frame it is still a finite theory, which
means that it can be renormalized just by field redefinitions, with no
redefinitions of parameters.

Calculating one-loop diagrams with four, six and eight external legs 
\[
\includegraphics[width=4.5truein,height=1truein]{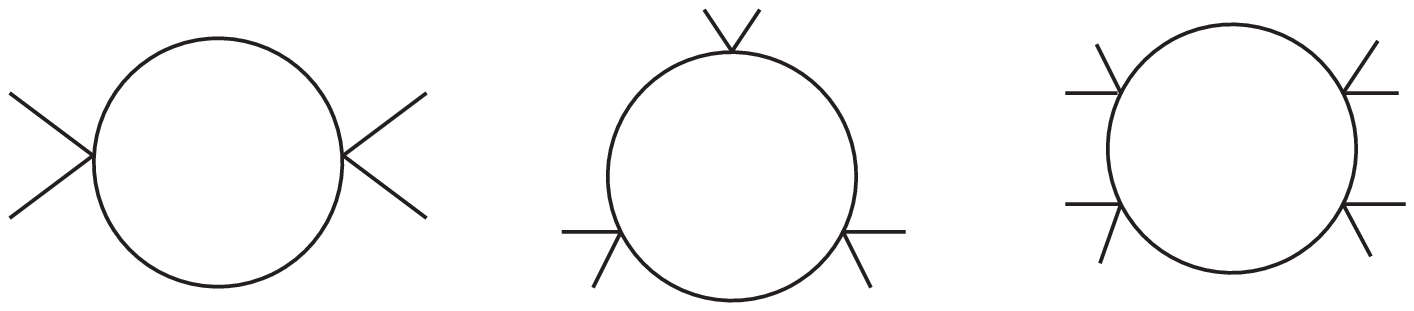} 
\]
we find the renormalized action 
\begin{eqnarray}
&&S(\varphi )=S_{c}(\varphi )+\frac{\hbar \lambda ^{4}}{512\pi
^{2}\varepsilon }\int \varphi ^{2}\left\{ 4(\Box \varphi )^{2}+\lambda
^{2}\varphi (\partial _{\mu }\varphi )^{2}\left[ 4(\Box \varphi )+\lambda
^{2}\varphi (\partial _{\nu }\varphi )^{2}\right] \right.  \nonumber \\
&&\qquad +\left. 2a\lambda ^{2}\varphi ^{2}(\Box \varphi )^{2}+b\lambda
^{4}\varphi ^{3}(\Box \varphi )(\partial _{\mu }\varphi )^{2}+c\lambda
^{4}\varphi ^{4}(\Box \varphi )^{2}\right\} +\hbar \mathcal{O}(\lambda
^{10})+\mathcal{O}(\hbar ^{2}),  \label{caguo}
\end{eqnarray}
where $a$, $b$ and $c$ are constants that we do not need to work out. The
theory is indeed finite up to the order $\mathcal{O}(\lambda ^{8})$
included, because the divergent terms of the first line cancel out using the
field equations, while those of the second line become $\mathcal{O}(\lambda
^{10})$. Observe that the cancelation is nontrivial, and occurs only because
the terms appearing on the first line of (\ref{caguo}) have coefficients
that are related in a way to make that happen. We find 
\[
S(\varphi )=S_{c}(\varphi _{\lambda }(\varphi ))+\hbar \mathcal{O}(\lambda
^{10})+\mathcal{O}(\hbar ^{2}), 
\]
where the field redefinition reads 
\begin{eqnarray}
\varphi _{\lambda }(\varphi ) &=&\varphi -\frac{\hbar \lambda ^{4}\varphi
^{2}}{512\pi ^{2}\varepsilon }\left[ 4(\Box \varphi )+2\lambda ^{2}\varphi
(\partial _{\mu }\varphi )^{2}+2(a-1)\lambda ^{2}\varphi ^{2}(\Box \varphi
)\right.  \nonumber \\
&&+\left. (c-a+1)\lambda ^{4}\varphi ^{4}(\Box \varphi )+(b-a)\lambda
^{4}\varphi ^{3}(\partial _{\mu }\varphi )^{2}\right] +\hbar \mathcal{O}%
(\lambda ^{10})+\mathcal{O}(\hbar ^{2}).  \label{yuge}
\end{eqnarray}
Combining the classical change of variables (\ref{classchv}) and the
renormalized one $\varphi _{\lambda }(\varphi )$, we get, up to $\hbar 
\mathcal{O}(\lambda ^{10})$ and $\mathcal{O}(\hbar ^{2})$, 
\[
S(\varphi )=S_{c}(\varphi _{\lambda }(\varphi ))=\frac{1}{2}\int \mathrm{d}%
^{D}x\hspace{0.01in}(\partial _{\mu }\varphi _{c}^{\prime }(\varphi
_{\lambda }(\varphi )))^{2}=S_{c}^{\prime }(\varphi _{c}^{\prime }(\varphi
_{\lambda }(\varphi ))). 
\]
The theory remains equivalent to a free massless field after
renormalization. Despite the unnecessary complications introduced by the
change of variables, the physics remains the same. Observe that the
renormalized field redefinition $\varphi _{c}^{\prime }(\varphi _{\lambda
}(\varphi ))$ is not derivative-independent anymore.

Before the change of variables, the relation between bare fields $\varphi _{%
\mathrm{B}}$ and renormalized fields $\varphi $ is $\varphi _{\mathrm{B}%
}=\varphi _{\lambda }(\varphi )$. After the change of variables the relation
between $\varphi _{\mathrm{B}}^{\prime }$ and $\varphi ^{\prime }$ is just $%
\varphi _{\mathrm{B}}^{\prime }=\varphi ^{\prime }$, because the theory is
manifestly free. Thus, the bare and renormalized changes of variables are 
\[
\varphi _{\mathrm{B}}^{\prime }=\varphi _{c}^{\prime }(\varphi _{\mathrm{B}%
}),\qquad \varphi ^{\prime }=\varphi _{c}^{\prime }(\varphi _{\lambda
}(\varphi )), 
\]
respectively.

\paragraph{Example 2\newline
}

Now we want to check the first line of (\ref{yuge}) using the method of
section 9. Instead of calculating diagrams in the variables $\varphi $, we
apply the change of variables to the renormalized theory written in the
variables $\varphi ^{\prime }$. There the theory is free, so we just need to
pay attention to the composite-field sector.

Expanding and inverting the classical change of variables (\ref{classchv}),
write 
\[
\varphi _{c}=\varphi ^{\prime }-\frac{\lambda ^{2}}{2}\frac{\varphi ^{\prime 
\hspace{0.01in}3}}{3!}+\frac{13\lambda ^{4}}{4}\frac{\varphi ^{\prime 
\hspace{0.01in}5}}{5!}+\mathcal{O}(\lambda ^{6}). 
\]
We have moved the subscript $c$ from $\varphi ^{\prime }$ to $\varphi $
since now we are making the transformation in the opposite direction. The
change of variables is thus expressed by the source redefinitions 
\begin{equation}
L_{3}^{\prime }=L_{3}-\frac{\lambda ^{2}}{2}J,\qquad L_{5}^{\prime }=L_{5}+%
\frac{13\lambda ^{4}}{4}J,  \label{hu}
\end{equation}
and so on, where $L_{i}^{\prime }$ is the source coupled to the composite
field $\varphi ^{\prime \hspace{0.01in}i}/i!\hspace{0.01in}$. Working at one
loop in the primed variable frame, the renormalized extended action $%
S_{L}^{\prime }$ is only made of the counterterms 
\begin{equation}
\frac{\hbar }{\varepsilon }\sum_{ij}r_{ij}\int \varphi ^{\prime \hspace{%
0.01in}i+j-4}L_{i}^{\prime }L_{j}^{\prime }=\frac{\hbar }{\varepsilon }\int
\varphi ^{\prime \hspace{0.01in}2}\left( r_{33}L_{3}^{\prime \hspace{0.01in}%
2}+r_{35}\varphi ^{\prime \hspace{0.01in}2}L_{3}^{\prime }L_{5}^{\prime
}+r_{55}\varphi ^{\prime \hspace{0.01in}4}L_{5}^{\prime \hspace{0.01in}%
2}\right) +\cdots  \label{count}
\end{equation}
where are $r_{ij}$ numerical constants. We will see that to check the first
line of (\ref{yuge}) it is sufficient to calculate $r_{33}$, which is given
by the diagram 
\[
\includegraphics[width=3truein,height=.9truein]{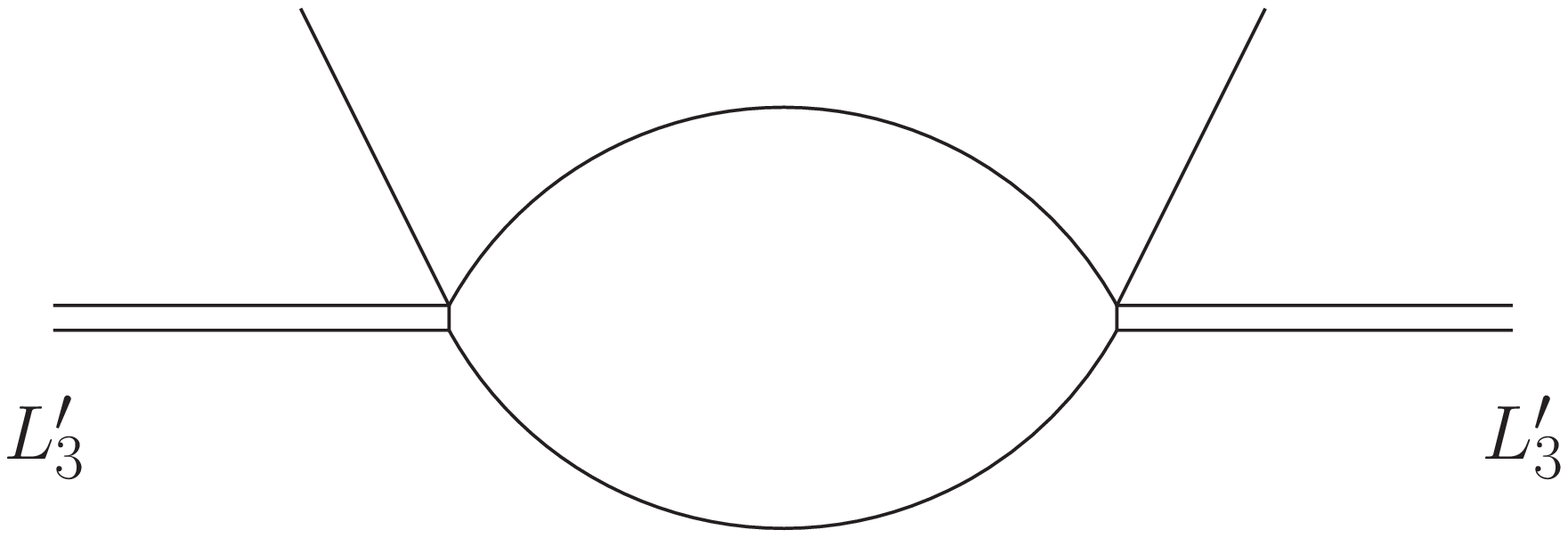} 
\]
and the one obtained exchanging the $L_{3}^{\prime }$-legs. We find $%
r_{33}=1/(32\pi ^{2})$.

Because of (\ref{count}), applying (\ref{hu}) we get an unprimed functional
integral that is written in some unconventional form. We can set $L_{i}=0$
now, since we do not need these sources anymore. The $J$-dependence in the
exponent of the $Z$-integrand reads 
\begin{equation}
\int J\left[ \varphi _{c}-\frac{\hbar \lambda ^{4}}{\varepsilon }\varphi
_{c}^{2}J\left( \frac{r_{33}}{4}+\frac{r_{33}-39r_{35}}{24}\lambda
^{2}\varphi _{c}^{2}\right) +\mathcal{O}(\lambda ^{8})\right] \equiv \int
J\left( \varphi _{c}+U(\varphi _{c},J)\right) .  \label{bosco}
\end{equation}
Dropping the subscript $c$ and making the change of integration variables $%
\varphi \rightarrow \varphi -U$, the term (\ref{bosco}) turns into $\int
J\varphi +\mathcal{O}(\lambda ^{8})$, but we get also contributions 
\[
-\int \frac{\delta S_{c}}{\delta \varphi }U+\mathcal{O}(\lambda ^{8})=\frac{%
\hbar \lambda ^{4}}{\varepsilon }\int \frac{\delta S_{c}}{\delta \varphi }%
\varphi ^{2}J\left( \frac{r_{33}}{4}+\frac{r_{33}-39r_{35}}{24}\lambda
^{2}\varphi ^{2}\right) +\mathcal{O}(\lambda ^{8}) 
\]
from the action. The integral is still written in an unconventional form,
and the $J$-dependence in the exponent of the $Z$-integrand becomes 
\[
\int J\left[ \varphi -\frac{\hbar \lambda ^{4}}{\varepsilon }\frac{\delta
S_{c}}{\delta \varphi }\varphi ^{2}\left( \frac{r_{33}}{4}+\frac{%
r_{33}-39r_{35}}{24}\lambda ^{2}\varphi ^{2}\right) +\mathcal{O}(\lambda
^{8})\right] . 
\]
The further change of variables 
\[
\varphi \rightarrow \varphi +\frac{\hbar \lambda ^{4}}{\varepsilon }\frac{%
\delta S_{c}}{\delta \varphi }\varphi ^{2}\left( \frac{1}{128\pi ^{2}}+\frac{%
r_{33}-39r_{35}}{24}\lambda ^{2}\varphi ^{2}\right) +\mathcal{O}(\lambda
^{8}) 
\]
finally takes us to the conventional form, up to the desired order. The
field renormalization in the unprimed variables is obtained composing the
changes of integration variables made so far and setting $J=0$. We conclude
that 
\[
\varphi _{\lambda }(\varphi )=\varphi +\frac{\hbar \lambda ^{4}}{128\pi
^{2}\varepsilon }\frac{\delta S_{c}}{\delta \varphi }\varphi ^{2}-\frac{%
\hbar \lambda ^{6}}{\varepsilon }\frac{r_{33}-39r_{35}}{24}(\Box \varphi
)\varphi ^{4}+\mathcal{O}(\lambda ^{8}), 
\]
in agreement with (\ref{yuge}).

\paragraph{Example 3\newline
}

It is instructive to consider linear changes of field variables in a theory
where the fields are renormalized multiplicatively. We can restrict the set
of composite fields to the elementary field itself, coupled with the source $%
L_{1}$, and the identity. The BR change of variables reads 
\begin{equation}
\varphi _{\mathrm{B}}=Z_{\varphi }^{1/2}\varphi ,\qquad \lambda _{\mathrm{B}%
}=\lambda _{\mathrm{B}}(\lambda ,\mu ),\qquad L_{0\mathrm{B}}=L_{0},\qquad
L_{1\mathrm{B}}=Z_{\varphi }^{-1/2}(L_{1}+J)-J,\qquad J_{\mathrm{B}}=J.
\label{era}
\end{equation}
Now, consider the bare and renormalized changes of variables 
\[
\varphi _{\mathrm{B}}^{\prime }=b_{0\mathrm{B}}+(1+b_{1\mathrm{B}})\varphi _{%
\mathrm{B}},\qquad \varphi ^{\prime }=b_{0}+(1+b_{1})\varphi . 
\]
The bare change of variables is implemented by 
\[
L_{0\mathrm{B}}^{\prime }=L_{0\mathrm{B}}-b_{0\mathrm{B}}J_{\mathrm{B}%
},\qquad L_{1\mathrm{B}}^{\prime }=L_{1\mathrm{B}}-b_{1\mathrm{B}}J_{\mathrm{%
B}}, 
\]
while the renormalized one is given by the same formula with $\mathrm{B}$s
suppressed. Closing the scheme (\ref{schiena}) we find $J_{\mathrm{B}%
}^{\prime }=J^{\prime }$ and 
\[
L_{0\mathrm{B}}^{\prime }=L_{0}^{\prime }+(b_{0}-b_{0\mathrm{B}})J^{\prime
},\qquad L_{1\mathrm{B}}^{\prime }=Z_{\varphi }^{-1/2}L_{1}^{\prime
}+[Z_{\varphi }^{-1/2}(1+b_{1})-1-b_{1\mathrm{B}}]J^{\prime }. 
\]
These relations, together with 
\[
\varphi _{\mathrm{B}}^{\prime }=b_{0\mathrm{B}}+Z_{\varphi }^{1/2}(1+b_{1%
\mathrm{B}})\frac{\varphi ^{\prime }-b_{0}}{1+b_{1}}, 
\]
give the renormalization in the new variables. We are free to choose
different relations between the bare and renormalized $b$s. Doing this we
obtain equivalent ways to describe the renormalization. For example, if we
choose 
\[
b_{0}=b_{0\mathrm{B}},\qquad b_{1\mathrm{B}}=Z_{\varphi }^{-1/2}(1+b_{1})-1, 
\]
we find that the field does not renormalize in the new variables: $\varphi _{%
\mathrm{B}}^{\prime }=\varphi ^{\prime }$. This is just because we have
rescaled $\varphi $ by a new parameter $1+b_{1}$ and transferred the
renormalization on that parameter.

\paragraph{Example 4\newline
}

The simplest nonlinear change of variables involves a quadratic term $%
\varphi ^{2}$. Thus, we study a free massless scalar field in the presence
of the composite field $\varphi ^{2}/2$.

The renormalized generating functional is 
\begin{equation}
Z(J,L)=\mathrm{e}^{W(J,L)}=\int [\mathrm{d}\varphi ]\exp \left( -\frac{1}{2}%
\int \left\{ (\partial _{\mu }\varphi )^{2}-L_{2}\varphi ^{2}-\frac{\mu
^{-\varepsilon }}{a}\left( 1+a\delta _{a}\right) L_{2}^{2}\right\} +\int
J\varphi \right) ,  \label{eW}
\end{equation}
where $\delta _{a}=-(16\pi ^{2}\varepsilon )^{-1}$ in dimensional
regularization. The functional integral is easy to work out, since it is
Gaussian. The source $L_{2}$ plays the role of (minus) a spacetime dependent
squared mass, so we obtain 
\begin{equation}
W(J,L)=\frac{1}{2}\int \left\{ J\frac{1}{-\Box -L_{2}}J+\mu ^{-\varepsilon
}\left( \frac{1}{a}+\delta _{a}\right) L_{2}^{2}\right\} -\frac{1}{2}\text{tr%
}\ln (-\Box -L_{2}).  \label{Gaussia}
\end{equation}

Let us find the secret identity satisfied by $\delta W/\delta L_{2}$. From (%
\ref{eW}) we get 
\begin{equation}
\frac{\delta W}{\delta L_{2}}=\frac{1}{2}\frac{\delta ^{2}W}{\delta J^{2}}+%
\frac{1}{2}\left( \frac{\delta W}{\delta J}\right) ^{2}+\mu ^{-\varepsilon
}\left( \frac{1}{a}+\delta _{a}\right) L_{2}.  \label{aiop}
\end{equation}
Working out the derivatives of $W$, given by (\ref{Gaussia}), it is easy to
check this identity explicitly. The last term on the right-hand side of (\ref
{aiop}) compensates for the divergence due to the $J$-derivatives at
coinciding points.

\paragraph{Example 5\newline
}

Now we study the change of variables $L_{2}=L_{2}^{\prime }+bJ$ in the
previous example, to the lowest order in $b$. We also derive the secret
identity that ensures the closure of (\ref{schiena}).

We introduce sources $L_{0}$ and $L_{1}$ for the identity and the elementary
field. Then the renormalized generating functional is (\ref{eW}) times $\exp
\int L_{0}$, with $J$ replaced by $J+L_{1}$. We work in the nonlinear
approach. There the bare quantities are equal to the renormalized ones apart
from 
\[
L_{0\mathrm{B}}=L_{0}+\frac{\mu ^{-\varepsilon }\delta _{a}}{2}L_{2}^{2}. 
\]
Since we do not need the parameter $a$ to reabsorb divergences, we work at $%
a=\infty $. The functional integral gives a $W(J,L)$ equal to (\ref{Gaussia}) plus $\int L_{0}$, with $J\rightarrow J+L_{1}$ and $%
a=\infty $. The bare generating functional is formally identical with $%
\delta _{a}\rightarrow 0$.

The bare redefinitions $L_{2\mathrm{B}}=L_{2\mathrm{B}}^{\prime }+b_{\mathrm{%
B}}J_{\mathrm{B}}$, $L_{1\mathrm{B}}=L_{1\mathrm{B}}^{\prime }$, $L_{0%
\mathrm{B}}=L_{0\mathrm{B}}^{\prime }$ are equivalent to the change of
variables $\varphi _{\mathrm{B}}^{\prime }=\varphi _{\mathrm{B}}+b_{\mathrm{B%
}}\varphi _{\mathrm{B}}^{2}/2$. At the renormalized level, the $L^{\prime
}=L^{\prime }(L,J)$-redefinition reads $L_{2}=L_{2}^{\prime }+bJ$, $%
L_{1}=L_{1}^{\prime }$, $L_{0}=L_{0}^{\prime }$. It can be studied using the
procedure of section \ref{rchov}. Doing so, we obtain new relations between
primed bare and renormalized quantities, namely 
\begin{equation}
L_{0\mathrm{B}}^{\prime }=L_{0}^{\prime }+\frac{\mu ^{-\varepsilon }\delta
_{a}}{2}L_{2}^{\prime }(L_{2}^{\prime }-2bL_{1}^{\prime }),\quad L_{1\mathrm{%
B}}^{\prime }=L_{1}^{\prime }-b\mu ^{-\varepsilon }\delta _{a}(\Box
L_{2}^{\prime }+L_{2}^{\prime \hspace{0.01in}2}),\quad L_{2\mathrm{B}%
}^{\prime }=L_{2}^{\prime }.  \label{brrr}
\end{equation}
Note that the field is non-renormalized also after the change of variables,
to the lowest order in $b$, which is why the BR relations (\ref{brrr}) are $%
J $-independent. Closing the scheme (\ref{schiena}) we find the alternative $%
L^{\prime }=L^{\prime }(L,J)$-redefinition 
\[
L_{0}^{\prime }=L_{0}+\frac{\mu ^{-\varepsilon }\delta _{a}}{\varepsilon }%
L_{2}(b_{\mathrm{B}}J+bL_{1}),\quad L_{1}^{\prime }=L_{1}+\frac{b\mu
^{-\varepsilon }\delta _{a}}{\varepsilon }(\Box L_{2}+L_{2}^{2}),\quad
L_{2}^{\prime }=L_{2}-b_{\mathrm{B}}J. 
\]
up to higher-orders in $b$. This redefinition differs from the one we made,
which was $L_{2}=L_{2}^{\prime }+bJ$, $L_{1}=L_{1}^{\prime }$, $%
L_{0}=L_{0}^{\prime }$. We can make the two coincide choosing $b_{\mathrm{B}%
}=b$, provided we can drop the divergent corrections. Such corrections have
no effect on $W$ provided the secret identity 
\[
\int (\Box L_{2}+L_{2}^{2})\frac{\delta W}{\delta L_{1}}+\int L_{2}(J+L_{1})%
\frac{\delta W}{\delta L_{0}}=0 
\]
holds. It is easy to check that it is indeed so, since 
\[
\frac{\delta W}{\delta L_{1}}=\int \frac{1}{-\Box -L_{2}}(J+L_{1}),\qquad 
\frac{\delta W}{\delta L_{0}}=1. 
\]

\section{Conclusions}

\setcounter{equation}{0}

In this paper we have developed a field-covariant approach to quantum field
theory, concentrating on the $Z$- and $W$-functionals. Because of the
intimate relation with composite fields $\mathcal{O}^{I}(\varphi )$,
ultimately a perturbative change of field variables can be expressed as a $J$%
-dependent redefinition of the sources $L_{I}$ coupled to the $\mathcal{O}%
^{I}(\varphi )$s. We have defined several approaches, useful for different
purposes, in particular a linear approach where all variable changes can be
described as linear redefinitions $L_{I}\rightarrow
(L_{J}-b_{J}J)(z^{-1})_{I}^{J}$, including the map relating bare and
renormalized quantities. The functionals $Z$ and $W$ behave as scalars. We
have also seen how to convert a functional integral written in an
unconventional form to the conventional form. Among the other things, this
operation allows us to relate the renormalization of variable-changes to the
renormalization of composite fields, and gives a simple method to derive the
renormalization of the theory in the new variables from the renormalization
of the theory in the old variables, without having to calculate diagrams
anew.

The formalism developed here allows us to abandon the description of
renormalization as a set of replacements, and view it as made of true
changes of field variables, combined with parameter-redefinitions. Instead
of jumping from a variable frame to another one, we can write down
identities relating the generating functionals before and after a change of
field variables. We regard these results as a first step to upgrade the
formalism of quantum field theory to a more evolved one. Other issues, such
as the effects of variable-changes on the $\Gamma $-functional, are treated
in separate works.

\end{document}